\documentclass[fleqn,usenatbib]{mnras}

\usepackage[T1]{fontenc}
\usepackage{ae,aecompl}

\usepackage{graphicx}	
\usepackage{amsmath}	
\usepackage{amssymb}	
\usepackage[utf8]{inputenc}
\usepackage{listings}
\usepackage{color}
\usepackage{tabularx}
\usepackage{arydshln}
\usepackage{hyperref}
\usepackage{comment}
\usepackage[table]{xcolor}  

\usepackage{newtxtext,newtxmath}

\newcommand{\uproman}[1]{\uppercase\expandafter{\romannumeral#1}}
\newcommand{\lowroman}[1]{\romannumeral#1\relax}
\definecolor{schrift}{gray}{.5}

\title[Sub-mm/mm optical constants from MIRO observations]{Sub-mm/mm optical properties of real protoplanetary matter derived from Rosetta/MIRO observations of comet 67P}

\author[J. B\"urger et al.]{
Johanna B\"urger$^{1}$\thanks{E-mail: j.buerger@tu-braunschweig.de},
Thilo Glißmann$^{1}$,
Anthony Lethuillier$^{1}$,
Dorothea Bischoff$^{1}$,
\newauthor
Bastian Gundlach$^{1}$,
Harald Mutschke$^{2}$,
Sonja H\"ofer$^{3}$,
Sebastian Wolf$^{4}$
\newauthor
and J\"urgen Blum$^{1}$\\
$^{1}$Institut für Geophysik und extraterrestrische Physik, TU Braunschweig, Mendelssohnstr. 3, D-38106 Braunschweig, Germany \\
$^{2}$Astrophysikalisches Institut und
Universitäts-Sternwarte, Friedrich-Schiller-Universität Jena, Schillergässchen 2-3, D-07744 Jena, Germany\\
$^{3}$Leibniz-Institut für Photonische Technologien e.V., Albert-Einstein-Straße 9, D-07745 Jena, Germany\\
$^{4}$Institut für Theoretische Physik und Astrophysik, Christian-Albrechts-Universität zu Kiel, Leibnizstr. 15, Kiel, 24118 Germany\\
}

\date{Accepted XXX. Received YYY; in original form ZZZ}

\pubyear{2022}

\begin{document}
\label{firstpage}
\pagerange{\pageref{firstpage}--\pageref{lastpage}}
\maketitle

\begin{abstract}

Optical properties are required for the correct understanding and modelling of protoplanetary and debris discs. By assuming that comets are the most pristine bodies in the solar system, our goal is to derive optical constants of real protoplanetary material. We determine the complex index of refraction of the near-surface material of comet 67P/Churyumov-Gerasimenko by fitting the sub-millimetre/millimetre observations of the thermal emission of the comet's sub-surface made by the Microwave Instrument for the Rosetta Orbiter (MIRO) with synthetic temperatures derived from a thermophysical model and radiative-transfer models. According to the two major formation scenarios of comets, we model the sub-surface layers to consist of pebbles as well as of homogeneously packed dust grains. In the case of a homogeneous dusty surface material, we find a solution for the length-absorption coefficient of $\alpha \approx 0.22~\mathrm{cm^{-1}}$ for a wavelength of 1.594~mm and $\alpha \geq 3.84~\mathrm{cm^{-1}}$ for a wavelength of 0.533~mm and a constant thermal conductivity of 0.006~$\mathrm{Wm^{-1}K^{-1}}$. For the pebble scenario, we find for the pebbles and a wavelength of 1.594~mm a complex refractive index of $n = (1.074 - 1.256) + \mathrm{i} \, (2.580 - 7.431)\cdot 10^{-3}$ for pebble radii between 1~mm and 6~mm. Taking into account other constraints, our results point towards a pebble makeup of the cometary sub-surface with pebble radii between 3~mm and 6~mm. The derived real part of the refractive index is used to constrain the composition of the pebbles and their volume filling factor. The optical and physical properties are discussed in the context of protoplanetary and debris disc observations.

\end{abstract}

\begin{keywords}
Physical data and processes: opacity; comets: individual: 67P/Churyumov-Gerasimenko; ISM: dust: extinction ; planets and satellites: formation
\end{keywords}



\section{Introduction}
ESA's space mission Rosetta was the first mission that entered an orbit around and landed on a comet. The instruments on board the Rosetta spacecraft investigated its target comet 67P/Churyumov-Gerasimenko (hereafter 67P) for more than two years. The Microwave Instrument for the Rosetta Orbiter (MIRO) was located along with ten other instruments on-board the orbiter of the spacecraft. MIRO was a radiometer operating at millimetre and sub-millimetre wavelengths (hereafter mm/sub-mm) and was, thus, able to measure the thermal radiation emitted by the sub-surface regions of the nucleus \citep{Gulkis2007, Schloerb2015}. In principle, this method allows to probe the sub-surface temperature of the cometary surface, which is of major importance for understanding the physical processes driving cometary activity, i.e., the emission of dust and the outgassing of volatiles \citep[see, e.g.,][]{Gundlachetal2020}.

The deduction of the sub-surface temperatures from the received solar radiation, however, requires the application of a radiative transfer model, because the emitted photons have to travel through the sub-surface material where absorption and scattering occurs. The efficiency of these two effects depends on the complex refractive index, whose imaginary part causes absorption and whose real part causes reflection and diffraction.

In this work, we use the MIRO observations to derive the sub-mm/mm optical properties of the near-surface material of comet 67P. In contrast to \citet{Schloerb2015} and \citet{Macher}, who assumed an absorption coefficient for their calculations, we use a thermophysical model, which provides a temperature profile for the upper layers of the cometary nucleus. We then derive the complex index of refraction by fitting the temperature measurements of 67P's sub-surface made by MIRO with synthetic temperatures derived from the thermophysical and radiative-transfer model. 

Since comets are believed to be the most pristine objects of our solar system, the derived refractive index of cometary matter can be used to estimate the optical properties of real protoplanetary matter in the sub-mm/mm wavelength regime. These results could be directly applied to observations of young dusty objects in space by ALMA, e.g. protoplanetary discs, or debris discs, to yield reliable dust masses. In addition, the acquired optical properties can help to identify the materials that make up a cometary nucleus.

Hence, this manuscript is structured as follows. First, in Sect.~\ref{MIRO} a short description of the MIRO instrument is given along with an analysis of the MIRO data. The thermophysical modelling of the cometary sub-surface layers as well as the derivation of synthetic MIRO brightness temperatures are presented in Sect.~\ref{Sec:TemperatureSimulation}. The derivation of the sub-mm/mm optical properties is described in Sect.~\ref{Sec:OpticalProperties} and the results are discussed in Sect.~\ref{Sec:discussion}. Sect.~\ref{Sec:application} provides an overview on their application to disc observations. Finally, Sect.~\ref{Sec:Conclusion} summarises our main results.

\section{MIRO observations} \label{MIRO}

\subsection{The MIRO measurements on the Rosetta orbiter}

The MIRO instrument is a dual-frequency heterodyne radiometer \citep{Gulkis2007} located on-board the orbiter of the Rosetta spacecraft. Its main objectives were to measure the thermal radiation emitted by the sub-surface (down to depths on the order of 10 cm) of the comet's nucleus and to identify molecules of interest (mostly water and its isotopes) in the coma of 67P. MIRO received the signal through a 30-cm diameter antenna, which was then transmitted to two heterodyne receivers. These receivers operated at wavelengths of 0.533 mm (hereafter sub-mm) and 1.594 mm (hereafter mm), respectively. The corresponding frequencies were 562.8 GHz (sub-mm) and 188.2 GHz (mm), respectively \citep[see][for details]{Gulkis2007}.

The MIRO data used in this study is the antenna temperature averaged over 1 second, $T_\mathrm{A}$. This temperature is derived from the power $P$ in Watts received by the telescope through

\begin{equation}
T_\mathrm{A} = \frac{P}{k \, \mathrm{d}\nu},
\label{PowerTemp}
\end{equation}

with $k$ and $d\nu$ being Boltzmann's constant and the bandwidth of the sub-mm and mm receiver channels with values of $d\nu = 1~\mathrm{GHz}$ and $d\nu = 0.5~\mathrm{GHz}$, respectively. This relationship is used alongside observations of two calibration targets with known temperatures (18°C and –47°C in 2014) to retrieve the antenna temperatures for all times of observations. MIRO collected antenna temperatures every 50.5 ms and observed the calibration targets every 34 minutes. The data was then averaged over one second steps.

For this work, we followed the same procedure as in \citet{Choukroun_2015} to convert the antenna temperature to the brightness temperature, which corresponds to the temperature of a black body (emissivity~=~1) filling the telescope beam and providing the same observed power. The antenna temperature is related to the brightness temperature through \citep{UserManual}

\begin{equation}
T_\mathrm{A} = \frac{1}{k} \frac{h\nu}{e^{h\nu/kT_\mathrm{B}}-1},
\label{AntennaTemp}
\end{equation}

with $T_\mathrm{B}$, $\nu$ and $h$ being the brightness temperature, the frequency of the signal of interest and the Planck constant, respectively. We use Eq.~\ref{AntennaTemp} to derive $T_\mathrm{B}$ for all times of interest.

\subsection{Selection of a suitable MIRO data set} \label{Sec:Selection}
The MIRO data are available on the ESA Planetary Science Archive. In order to compare the brightness temperatures measured by MIRO with temperatures derived through the application of a thermophysical and radiative-transfer model, a suitable set of temperatures needs to be selected and then further filtered for optimal data quality. In the following, our approach as well as the parameters important for the filtering are presented.

We selected MIRO measurements that were taken when the orbiter was close to the surface of comet 67P in order to obtain the highest spatial resolution on the surface. This allowed us to focus our study on specific regions and areas of interest of the comet nucleus. We are particularly interested in the flat areas as this allows to use the least complex thermophysical model. Indeed, it was shown that surface roughness and the presence of shadows or cliffs, as well as self-heating effects need to be taken into account in non-flat regions \citep{Macher}. By choosing particularly flat areas, we avoid these effects and increase the reliability of our results.

We found such conditions just before equinox in March 2016 when the orbiter was as close as 11 km from the surface of 67P and observed the Imhotep region. This region is characterised by its flatness and location close to the equator on 67P's big lobe \citep{Auger}.

\begin{figure}
	\begin{center}
		\includegraphics[width = 1\columnwidth]{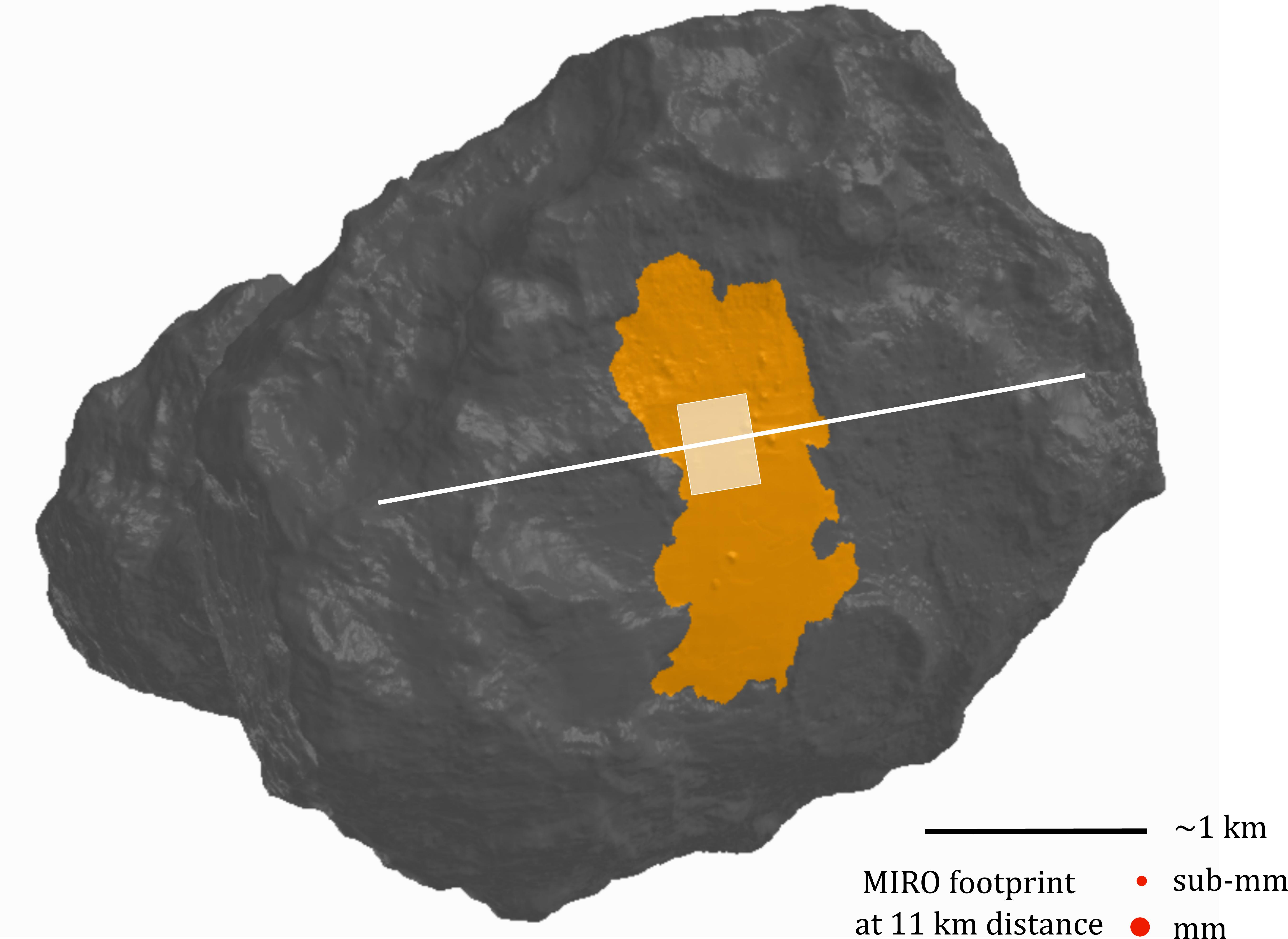} 
		\caption{The Imhotep region is located on the big lobe of comet 67P. We analysed MIRO measurements in Imhotep's sub-region 'a', which is coloured in orange. However, as there are parts that are further away from the equator (shown as a white line) and therefore illuminated differently, we select only measurements for which the centre of MIRO's beam was located in close proximity to the equator, as indicated by the white box.}
		\label{Fig:Region29}
	\end{center}
\end{figure}

Imhotep consists of four sub-regions 'a-d' \citep{Thomas}. We selected the temperatures measured in sub-region 'a', because it is the closest to the equator and the smoothest of all four sub-regions. Fig.~\ref{Fig:Region29} shows sub-region 'a' in orange, superimposed on the shape model of 67P.

\begin{figure}
    \begin{center}
		\includegraphics[width=1\columnwidth]{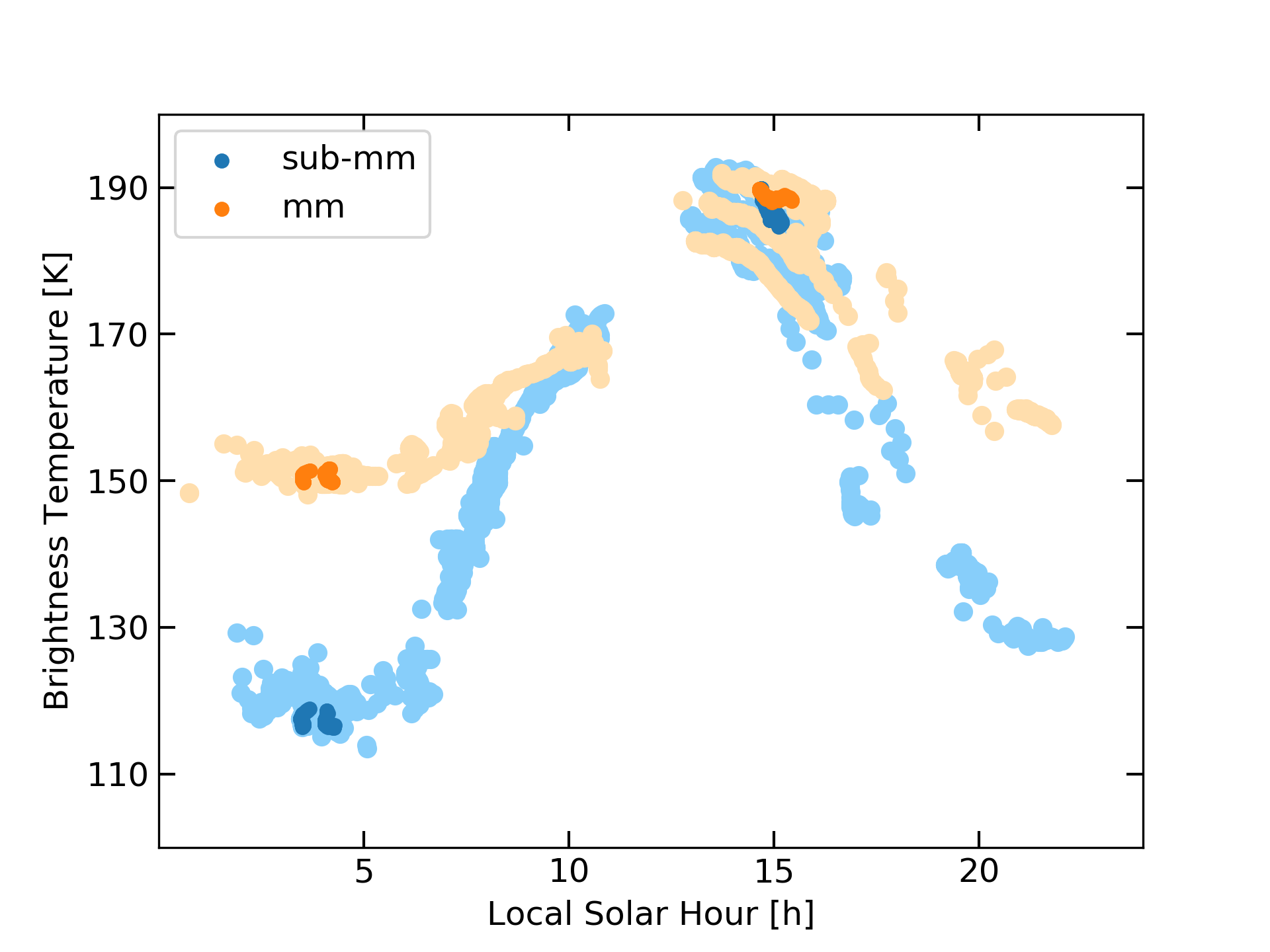}
	\end{center}	
	\caption[]{The brightness temperatures in Kelvin are plotted against the effective local solar hour, which is based on a 24~h day. 0~h corresponds to midnight and 12~h to noon. Shown are the brightness temperatures measured in Imhotep's sub-region 'a' in March 2016. Highlighted are the filtered temperatures, according to the white box in Fig. \ref{Fig:Region29}.}
	\label{Fig:Temp_curves_MIRO}
\end{figure}

Fig.~\ref{Fig:Temp_curves_MIRO} shows a diurnal temperature curve for Imhotep's sub-region 'a' with measurements made in March 2016. The brightness temperatures obtained for the sub-mm/mm channels are plotted against the effective local solar hour on 67P. The effective local solar hour is based on a 24 h day, where 0~h corresponds to midnight and 12~h to noon. These temperature data were filtered by emission angles smaller than 60\textdegree{}. The emission angle is defined as the angle between the surface normal and the MIRO boresight. For such restricted emission angles, the effect of signal polarisation has little to no effect on the brightness temperature. Furthermore, we selected only measurements that were made at distances smaller than 20 km.

As sub-region 'a' of Imhotep still covers parts of 67P that are not close to the equator and are therefore illuminated differently (see Fig.~\ref{Fig:Region29}), we further filtered by latitude and longitude, for which we selected ranges for the centre of MIRO's beam of +10\textdegree{} to -10\textdegree{} latitude and 125\textdegree{} to 135\textdegree{} longitude, respectively (see Fig.~\ref{Fig:Region29}, white box). This selected area is covered predominately by Imhotep's sub-region 'a'.

The selected temperatures are highlighted with darker colours in Fig.~\ref{Fig:Temp_curves_MIRO}. Three sets of temperatures were found for each of the two MIRO channels. One measurement was made on 19 March 2016 between 14:36~h and 15:24~h local solar hour, the other two data sets were recorded two days later on 21 March 2016 between 3:28~h and 3:39~h as well as between 4:00~h and 4:12~h local solar hour. Hereafter, the three sets of temperatures will be referred to as the day, night \uproman{1} and \uproman{2} cases, respectively. Although the night \uproman{2} measurements are later at night, the temperatures measured are similar due to a different emission angle, which is on average 40\textdegree{} in the night \uproman{1} case and 30.5\textdegree{} in the night \uproman{2} case. With a smaller emission angle, radiation arising from deeper layers is measured, which are warmer during night. Consequently, in the night \uproman{2} case, the upper layers could already be cooler than in the night \uproman{1} case, but MIRO could still measure similar temperatures.

We use the mean values of the brightness temperatures measured within the above given time interval for the different cases. The respective minimum and maximum brightness temperatures provide us together with the uncertainties of the antenna temperatures with an uncertainty estimate of the MIRO measurements. The measurement uncertainties of MIRO's antenna temperatures are $\pm 1.0$~K in the sub-mm channel and $\pm 0.4$~K in the mm channel \citep{UserManual}. We do not take into account the error caused by the widths of the emission angle and the latitude and longitude ranges on the cometary surface. To obtain the total error of the selected measurements, we added the squares of the MIRO random errors and the measurement dispersion and took the square root. To this calculated error, we added the calibration error of 1~K \citep[as shown in Table~1 of][]{Schloerb2015}. The MIRO beam efficiency is a known effect on the measurements, but this effect can only be calculated by using the full MIRO beam pattern to determine the local illumination. This usually is unpractical, as it requires longer computation time, but in our case for which the number of measurement times analysed is fairly small, this was done. Therefore, the illumination pattern used for the numerical thermal model takes into account the beam efficiency error. The selected mean brightness temperatures together with the corresponding uncertainties are presented in Table~\ref{Tab:TB_MIRO}.

\begin{table}
	\caption{Mean brightness temperatures $T_\mathrm{B}$ measured by MIRO for the three selected cases.}
	\centering
	\begin{tabular}{|l|l||c|}
		\cline{1-3}
		\multicolumn{2}{|l||}{Case} & $T_\mathrm{B}$ in K\\\hline\hline
		Day & mm & 188.7 $\pm$ 2.2 \\
		& sub-mm & 186.6 $\pm$ 4.2 \\
		\cline{1-3}
        Night \uproman{1} & mm & 150.6 $\pm$ 2.0 \\
		& sub-mm & 117.5 $\pm$ 2.7 \\
		\cline{1-3}
		Night \uproman{2} & mm & 150.6 $\pm$ 2.2 \\
		& sub-mm & 117.1 $\pm$ 2.9 \\
		\cline{1-3}
	\end{tabular}
	\label{Tab:TB_MIRO}
\end{table}

Table~\ref{Tab:keydata} lists 67P's heliocentric distance, the distance of the Rosetta orbiter to 67P's surface at that time as well as the key dates of the selected temperature measurements. MIRO's footprint, referring to the beam size on 67P's surface, is calculated to be about 24 m (sub-mm) and 74 m (mm) in diameter, respectively.

\begin{table*}
	\caption[Key data]{Key data for the observed region and time on comet 67P used in this study. lsh stands for local solar hour which is based on a 24 h day.}
	\centering
	\begin{tabular}{|l||l|}
		\cline{1-2}
		Measurements with MIRO & day: 19 March 2016 at 14:36 - 15:24 (lsh)\\
		& night \uproman{1}: 21 March 2016 at 3:28 - 3:39 (lsh); night \uproman{2}: 21 March 2016 at 4:00 - 4:12 (lsh)\\
		Heliocentric distance & 2.61 AU (day case); 2.63 AU (night cases)\\
		Distance spacecraft to 67P's surface & $\approx$ 11 km \\
		MIRO footprint (diameter) & 24 m (sub-mm) and 74 m (mm)\\
		Emission angle & 46 \textdegree{} (day case); 40 \textdegree{} (night \uproman{1} case); 30.5 \textdegree{} (night \uproman{2} case)\\
		\cline{1-2}
	\end{tabular}
	\label{Tab:keydata}
\end{table*}

\subsection{Daytime shadowing effects} \label{Sec:shadowing}
During the daytime, the brightness temperatures measured in both channels are quite comparable (see Fig.~\ref{Fig:Temp_curves_MIRO}). However, for normal matter, we would expect an increase in the absorption efficiency with decreasing wavelength and, thus, higher temperatures in the sub-mm channel, because the upper layers should be warmer due to a stronger heating by the Sun. During the nighttime, we see the expected difference between the two channels -- the sub-mm channel measures lower temperatures compared to the mm channel, because the upper layers are colder than the deeper ones. The problem with rather high absorption is that during the daytime, shadowing effects are non-negligible, because the cometary surface possesses a finite roughness, even if the region observed is relatively flat. This  results in MIRO measurements being a mixture of warmer (in sunlight) and cooler (in shadow) temperatures within the first few millimetres/centimetres, which predominantly would affect the measurements in the sub-mm channel, while the temperatures measured in the mm channel would be less affected, because deeper layers contribute more to the total flux. 
To give a quantitative example: The thermophysical model (see Sect.~\ref{sec:tpm}) predicts a surface temperature of about 220~K at 15~h local solar hour, compared to a measured brightness temperature of about 185~K in the sub-mm channel. This corresponds to a flux deficit of about 15\%. The measured brightness temperature in the sub-mm channel can be explained by a linear combination of warmer (about 220~K) and colder (about 120~K) temperatures, the latter due to shadowing and surface roughness. Thus, the MIRO daytime sub-mm temperatures (and possibly also the daytime mm temperatures) must be treated as lower limits. This will be taken into account in the data analysis.

Daytime shadowing effects on radiometric measurements have been discussed before. \citet{Grott} and \citet{ogawaetal2019} experienced the same problem with daytime observations when analysing the brightness temperatures of asteroid Ryugu, measured with the infrared radiometer MARA on board MASCOT. Stating that the daytime data are affected by topography and surface roughness, they only used the nighttime data to fit thermophysical models.

\section{Thermophysical modelling of the cometary sub-surface and the derivation of synthetic MIRO temperatures} \label{Sec:TemperatureSimulation}

For the comparison with MIRO observations, synthetic temperatures have to be produced. Therefore, a one-dimensional thermophysical model was chosen to represent the sub-surface of comet 67P. The output of the thermophysical model is a temperature-depth profile that serves as input for a radiative-transfer model, whose output is, in turn, a synthetic brightness temperature that can be compared to the MIRO measurements.

In the following Subsections, the general physical makeup of a cometary nucleus, the thermophysical model as well as the radiative-transfer model will be described. 

\subsection{Physical structure of cometary nuclei} \label{sec:formation}
To allow for the most reasonable morphology and makeup of the cometary sub-surface dust layer, we considered the two major formation scenarios for planetesimals \citep{Blum2014,Blum2017,Blum2018,Weissmanetal2020} and assume that present-day comets still possess their pristine makeup from their formation era. 

\begin{figure*}
	\begin{center}
		\includegraphics[width=1\columnwidth]{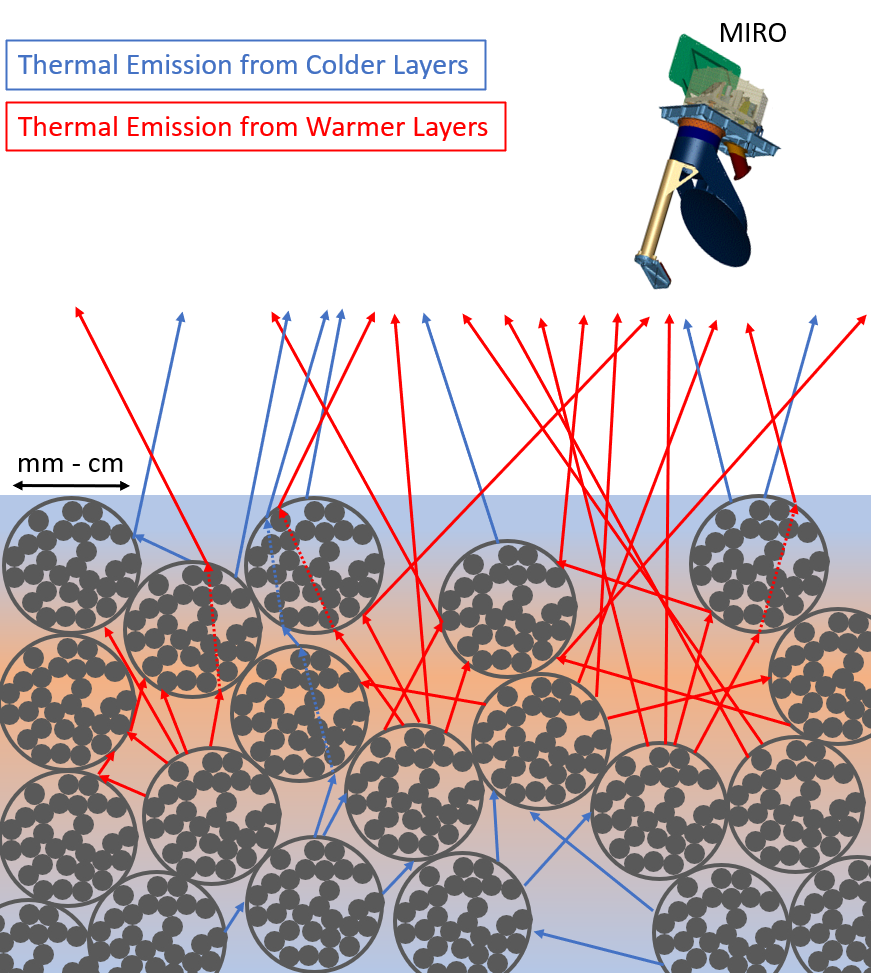}
		\includegraphics[width=1\columnwidth]{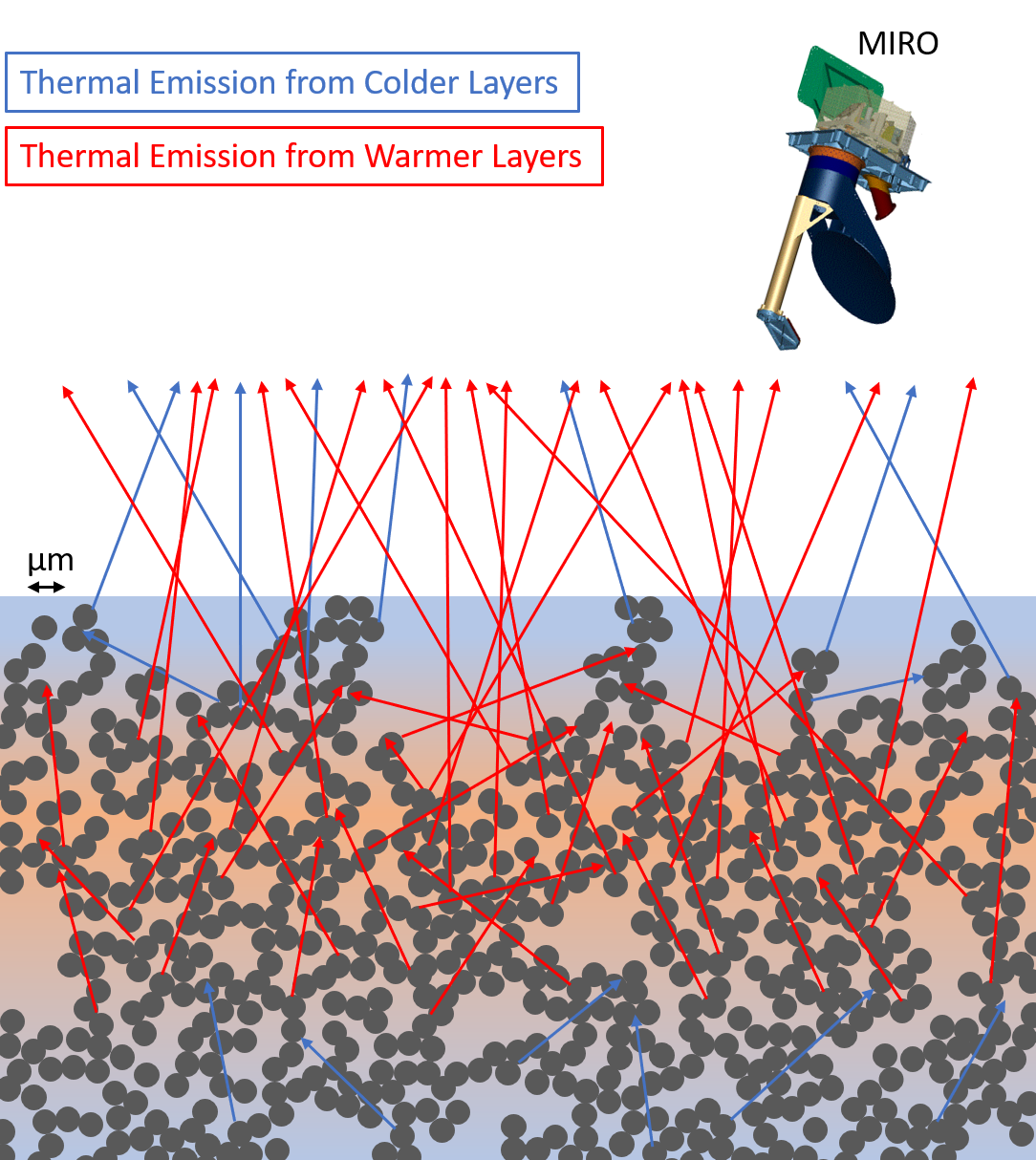}
		\caption[]{Sketches of the two models of the makeup of the cometary subsurface regions, including the radiative transfer of thermal emission inside and out of the nucleus. In the top right corner of each panel, the MIRO instrument is shown. Only rays emitted from the comet in this direction can be detected by MIRO. The colours in the background mark a night case, in which a warmer layer is located between two cold layers (see Figure \ref{Fig:TemperatureProfiles_example} for examples). \textbf{Left:} The pebble model with typical pebble sizes in the mm-cm range. The arrows mark thermal radiation that can be emitted (arrow starts), absorbed (arrow ends) or refracted at the pebble surface (dotted lines inside pebbles). \textbf{Right:} The no-pebble model as a loose assemblage of $\mathrm{\mu m}$-sized dust grains. The arrows mark thermal radiation that can either be emitted (arrow starts) or absorbed (arrow ends). Due to the large length contrast between the MIRO wavelengths and the dust-particle sizes, scattering or refraction does not occur.}
	\label{Fig:TheTwoModels}
	\end{center}
\end{figure*}

In the first formation model (see Figure \ref{Fig:TheTwoModels}, left), the cometary nucleus is composed of mm- to cm-sized porous aggregates (a.k.a. pebbles) consisting of (sub-)$\mathrm{\mu m}$-sized dust and ice grains. The pebbles are the result of the protoplanetary dust-coagulation process and are distinct and abundant, because they mark the transition between sticking and bouncing, the so-called bouncing barrier \citep{Zsom2010,Lorek2018}. Hydrodynamic concentration processes lead to the gentle gravitational collapse of clouds of pebbles \citep{JohansenEtAl2007} and to the formation of planetesimals with a hierarchical structure and porosity on two different length-scales, intra-pebble and inter-pebble porosity \citep{SkorovBlum2012}. The consequence of this particular makeup is a very low solid-state heat conductivity, due to the rather small number of grain contacts between the pebbles. On the other hand, the porous pebble packing produces rather large void spaces that can allow for an efficient radiative energy transport \citep{Bischoff2021}.

The alternative model for the growth of planetesimals, namely by successive sticking collisions between the growing building blocks \citep{Davidsson2016}, predicts a homogeneous, but porous makeup of the surface regions that also consist of (sub-)$\mathrm{\mu m}$-sized dust and ice grains \citep{Blum2018} with no macroscopic voids (see Figure \ref{Fig:TheTwoModels}, right). In comparison to the pebble-pile makeup, this formation scenario results in a larger solid-state heat conductivity and a negligible contribution by radiative heat transfer, due to the microscopic scale of the voids  \citep{Bischoff2021}.

One fundamental physical difference between the two makeups of the cometary nucleus that is of relevance for the interpretation of the MIRO data is the thermal conductivity. The pebble model predicts extremely low values for the solid-state heat conductivity, due to the hierarchical makeup and the small contact area between the pebbles, but possesses a significant contribution of radiative heat exchange, due to the large mean free path for photons in the void spaces between the pebbles. Thus, the total heat conductivity can be expressed as
\begin{equation}
    \lambda_\mathrm{Pebble} = \lambda_{0,\mathrm{Pebble}} + a(r) ~ T_z^3 \,
    \label{eq:pebblehc}
\end{equation}
with $\lambda_{0,\mathrm{Pebble}}$, $a(r)$ and $T_z$ being the solid-state heat conductivity, a temperature-independent proportionality factor that linearly depends on the pebble radius $r$ and the pebble temperature at depth $z$, respectively \citep[see][for details]{Bischoff2021}.

For the non-pebble model, on the other hand, radiative heat transport is negligible for all expected temperatures, due to the small ($\mathrm{\mu m}$-sized) inter-grain void spaces so that the heat conductivity is solely determined by the solid-state contacts among the constituent grains. Thus, we can write the heat conductivity as
\begin{equation}
    \lambda_\mathrm{No-pebble} = \lambda_{0,\mathrm{No-pebble}} \,
        \label{eq:nopebblehc}
\end{equation}
with $\lambda_{0,\mathrm{No-pebble}}$ being the solid-state heat conductivity of the non-hierarchical model.

For simplicity, we assume that both, $\lambda_{0,\mathrm{Pebble}}$ and $\lambda_{0,\mathrm{No-pebble}}$, are independent of temperature and that the relation 
\begin{equation}
    \lambda_{0,\mathrm{Pebble}} \ll \lambda_{0,\mathrm{No-pebble}}
\end{equation}
holds for all pebble sizes of relevance \citep[][]{Bischoff2021}. The validity of Eq. \ref{eq:pebblehc} is limited to pebble sizes of $10 \, \mathrm{\mu m} \lesssim r \lesssim 1 \,  \mathrm{cm}$, because for too small pebble sizes the pebble-pebble contact model breaks down and for too large pebble sizes the underlying assumption of isothermal pebbles is no longer valid.

In essence, this dual approach is as general as it can be to distinguish between a temperature-independent (no-pebble case) and strongly temperature-dependent (pebble case) heat conductivity and is not only plausible as a result of comet formation, but may also be used for a distinction between a micro-porous and macro-porous morphology of the cometary subsurface layers that may have arisen by, e.g., fallback of dusty material ejected elsewhere on the nucleus. Thus, the choice of the Imhotep region (see Sect. \ref{Sec:Selection}) is a compromise between optimal observation conditions and primitiveness of the material. Imhotep might not be the most primordial region and is possibly covered by fallback material. However, this material can still be regarded as primitive in terms of the morphology of the material (pebbles vs. no-pebbles), because fallback occurs at such low velocities that pebbles and no-pebble chunks of material will most likely survive intact \citep[see the dust-aggregate collision model by][]{guettler2010}. Due to the expected loss of volatiles for fallback material, Imhotep even has the advantage that omission of latent-heat effects can be justified (see Sect. \ref{sec:tpm}).

\subsection{Thermophysical model} \label{sec:tpm}

We used a thermophysical model that solves the heat-transfer equation for a homogeneous medium, which is divided into several layers in depth \citep[see][for details]{Gundlachetal2020}. To simplify matters, we assumed desiccated sub-surface layers. Due to the relatively large heliocentric distance of $\sim 2.6$~AU during the MIRO measurements considered here, the cometary outgassing activity is rather low \citep{Combietal2020} so that the relative amount of energy converted into latent heat is low too. Ignoring outgassing effects makes the simulations also more reliable, because a quasi-static situation can be considered, without the complex treatment of the removal of overlaying dust layers \citep{Gundlachetal2020}. Treating only desiccated dust layers implies that latent-heat cooling, transport of water vapour, pressure build-up inside the surface layers and the ejection of dusty surface material does not have to be taken into account. Moreover, the mass density is constant in this case. In the model, the thermal conductivity depends strongly on the material void space dimension and on temperature.

For the simulations, we used the temperature-dependent heat capacity of the carbonaceous chondrite NWA 7309 \citep{Opeil2020}. Furthermore, for the thermal conductivity, we considered two different variants depending on the formation scenario. First, from previous studies \citep{GundlachBlum2012, Gundlachetal2020} we know that the pebble structure possesses relatively large void spaces in between the material. Consequently, thermal radiation in the pore spaces between the pebbles is an important factor that has to be considered. This implies that the thermal conductivity strongly depends on temperature. Hence, our calculations for pebble surfaces are based on the full description of the thermal conductivity \citep[see section A1.1 in][]{Gundlachetal2020}. Second, for the homogeneous surface, however, we can make the assumption to neglect the temperature dependency, because the pores between the grains are on the order of (sub-)micrometres and thus the radiative energy transport is extremely suppressed. To consider different homogeneous surfaces (grain size, packing structure, etc.), we decided to use a set of different, but individually constant thermal conductivity values.

For the solid material, we chose for simplicity in both models a temperature-independent thermal conductivity. This is not exactly what was found for NWA 7309 in \citet{Opeil2020}, but the thermal conductivity varies only by $\pm$ 10 \% in the range of 100 K to 200 K, which is a rather small effect.

\begin{figure}
	\begin{center}
		\includegraphics[width=1\columnwidth]{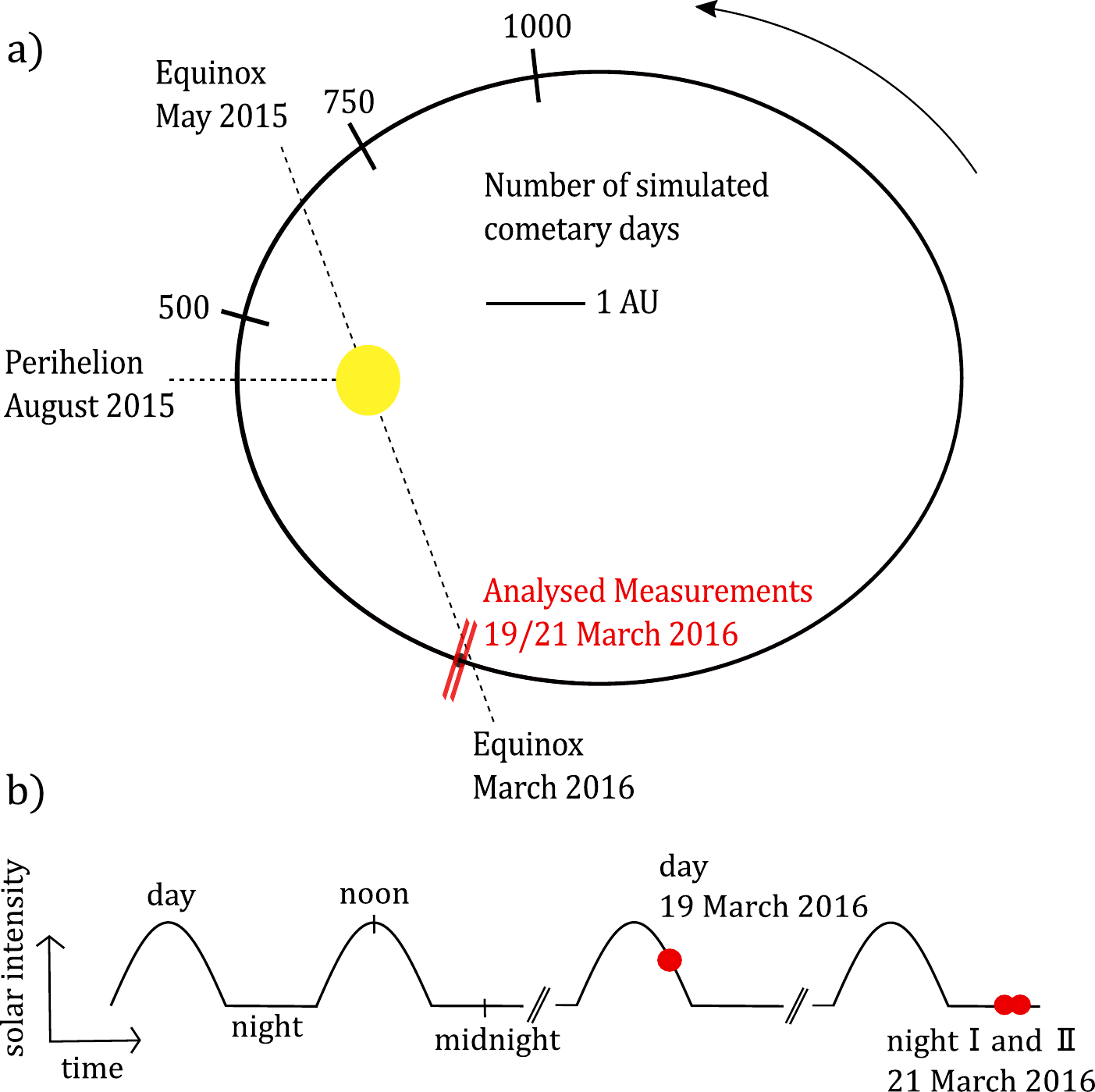}
		\caption[]{\textbf{a)} Comet 67P's trajectory around the Sun together with key dates. Locations of the analysed MIRO measurements are coloured in red. Drawn on the trajectory are the different numbers of cometary days simulated with the thermophysical model. The day number refers to the last measurement on 21 March 2016. 
		\textbf{b)} Visualisation of the illumination conditions in the simulation, i.e., solar intensity vs. time. The simulation starts at sunrise and runs through day and night. As the Sun's intensity is a sinusoidal function during daytime, the highest intensity is reached at noon. During night, the intensity is zero. To retrieve the temperature profile for the analysed MIRO data, the simulation is stopped at the respective local times for the day, night \uproman{1} and \uproman{2} cases, illustrated as red dots.}
	\label{Fig:SimulationNumberofCometaryDays}
	\end{center}
\end{figure}

At the start of the simulation, the temperature array is set to $50 \, \mathrm{K}$ at all depths. We tested different simulation settling times so that the comet surface layers are given enough time to adapt to the situation before the final observation dates are reached (19th and 21st of March, 2016; see Fig.~\ref{Fig:SimulationNumberofCometaryDays}). The tested simulation settling times are 500 days, 750 days and 1,000 days, respectively. Fig.~\ref{Fig:Temperatureprofiles_Night1} shows how in the night I case these timescales influence the resulting temperature-depth profiles. To realise these calculations, we used as input parameters the real heliocentric distance of the comet and the orientation of the analysed surface region to the Sun at every simulation day. In the end, we decided to use 1,000 days as simulation settling time throughout all thermophysical simulations. Thus, as Fig.~\ref{Fig:Temperatureprofiles_Night1} indicates, the accuracy of the simulations should be better than 1~K. The simulations then always end at the dates indicated in Table \ref{Tab:keydata}, but at different times (day, night I and night II). 

Our simulation runs were performed for cometary sub-surface dust layers either composed of equal-sized pebbles with the pebble radius as a free parameter or for a homogeneous surface material (see Sect. \ref{sec:formation}). In the case of a comet nucleus consisting of pebbles, we chose pebble radii of $r=1~\mathrm{mm}$, $r=3~\mathrm{mm}$, $r=5~\mathrm{mm}$, and $r=6~\mathrm{mm}$, respectively, slightly extending the pebble-size range predicted by \citet{Blum2017}. As for relatively large pebbles the heat transfer in the comet nucleus sub-surface is mostly dominated by inter-pebble radiation \citep[see Fig. 1 in][]{Bischoff2021}, the thermal conductivity is proportional to the mean-free path of the radiation between the pebbles, which, in turn, is proportional to the pebble radius. In the case of the homogeneous material (hereafter the no-pebble case), we chose temperature-independent heat conductivities between $\lambda=10^{-4} ~ \mathrm{W~m^{-1}~K^{-1}}$ and $\lambda=10^{-2} ~ \mathrm{W~m^{-1}~K^{-1}}$ to cover all possible cases.

We adapted the numerical resolution (i.e., the physical distance between the grid points, or the layer thickness) depending on the chosen scenario. For the pebble case, the resolution was set to one pebble radius, whereas for the no-pebble case, we used a resolution between $1 \, \mathrm{mm}$ and $5 \, \mathrm{mm}$. 

\begin{figure*}
	\begin{center}
		\includegraphics[width=0.8\textwidth]{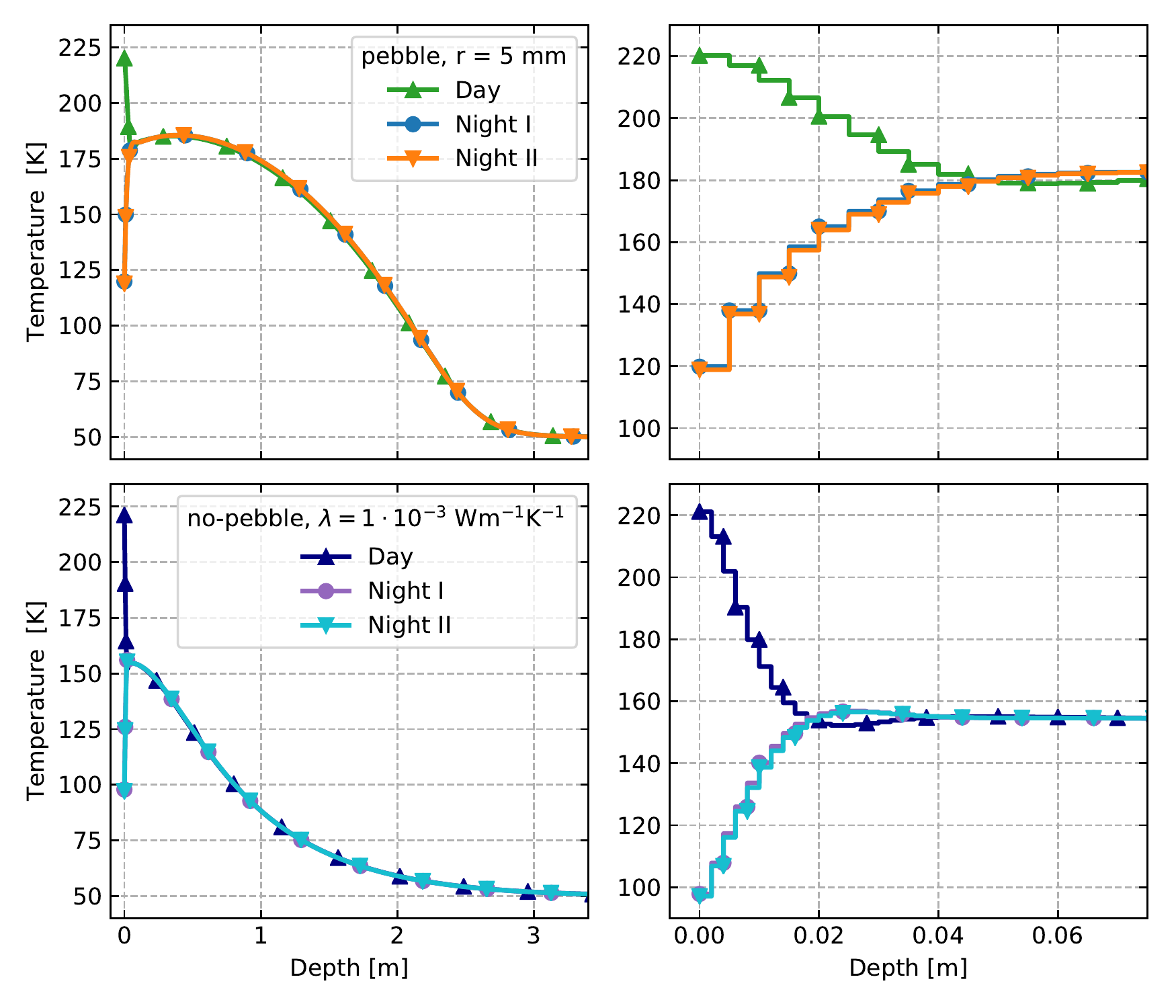}
		\caption{Two examples of derived temperature profiles. The left panels show the whole temperature-depth profile, whereas the right panels zoom into those sub-surface regions MIRO was sensitive to. All derived temperature profiles can be found in Appendix~\ref{AP:TP} (Fig.~\ref{Fig:TemperatureProfiles_pebble} and Fig.~\ref{Fig:TemperatureProfiles_nopebble}).}
		\label{Fig:TemperatureProfiles_example}
	\end{center}
\end{figure*}

Fig.~\ref{Fig:TemperatureProfiles_example} shows two examples of derived temperature profiles, one for the pebble case and one for the no-pebble case. The derived temperature profiles for the two night cases are very similar as they are close in time. For the day case, all temperature profiles, independent of pebble or no-pebble, show a surface temperature of around 220 K. For the night cases, smaller pebble sizes (pebble case) and smaller thermal conductivity values (no-pebble case) result in lower surface temperatures and a steeper increase of temperature within the first centimetres of the sub-surface. All resulting temperature profiles with 1,000 days settling time are shown in Fig.~\ref{Fig:TemperatureProfiles_pebble} for the pebble cases and in Fig.~\ref{Fig:TemperatureProfiles_nopebble} for the no-pebble cases.

\subsection{Radiative-transfer model and the derivation of synthetic MIRO temperatures} \label{Sec:rtm}
We apply two different radiative-transfer models for the no-pebble and the pebble case, which are presented in the following.

\subsubsection{Lambert-Beer law} \label{Sec:lb}
As we consider two wavelengths in the sub-mm/mm region and assume that the size of the solid constituents of the comet nucleus are much smaller than these wavelengths, scattering effects can be neglected so that we can apply an effective-medium model in the no-pebble case. 

\begin{figure}
	\begin{center}
		\includegraphics[width =0.5\columnwidth]{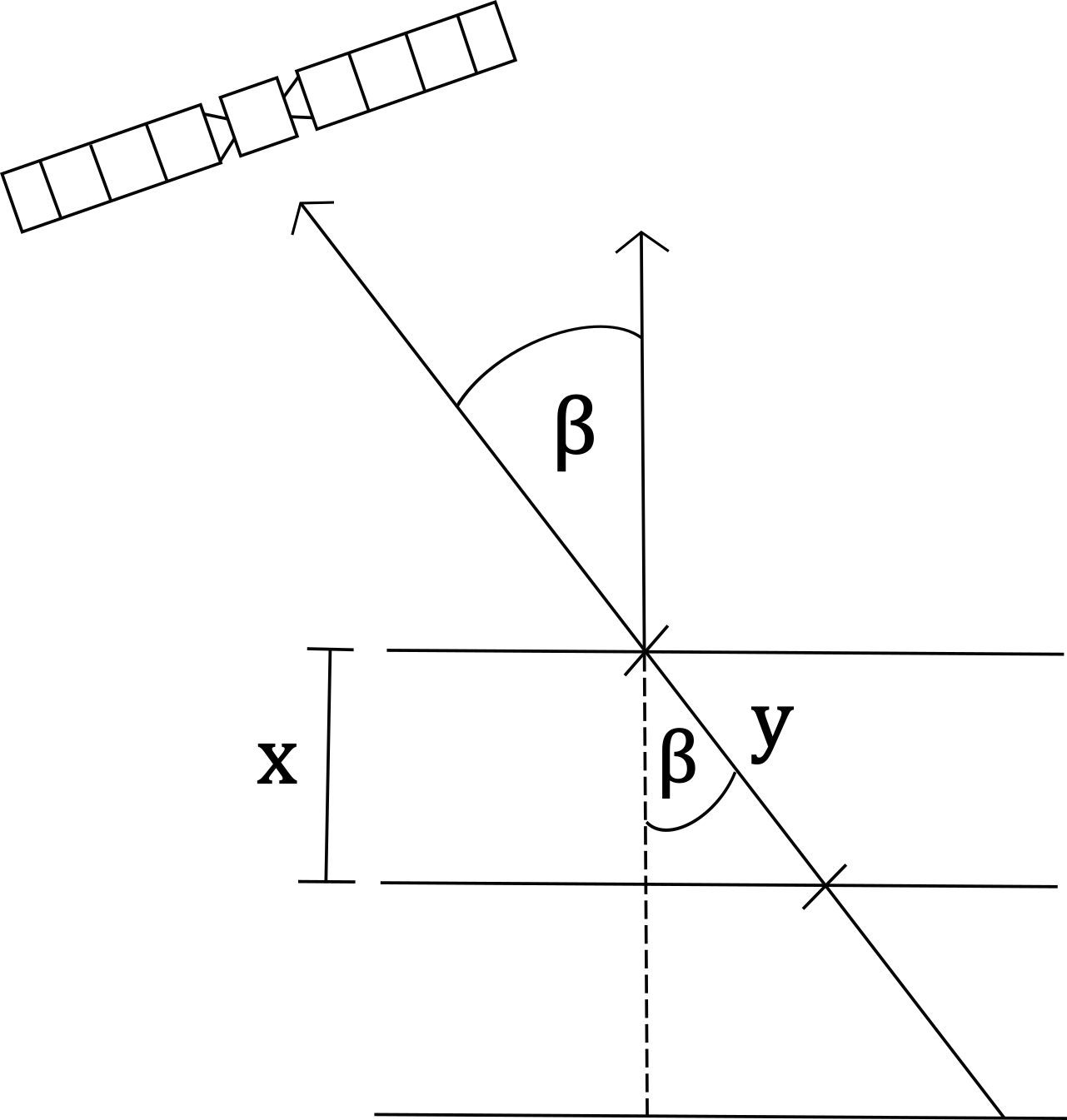}
		\caption{The emission angle $\beta$ is taken into account by replacing the depth $x$ of each layer by the optical path $y$.}
		\label{Fig:RadiativeTMEmissionAngle}
	\end{center}
\end{figure}

We derive the intensity $I_{\nu}$ received by MIRO by the following emission-absorption algorithm based on the Lambert-Beer law
\begin{equation}
I_{\nu}(T) = \int_{0}^{\infty} \alpha_{\nu} B_{\nu}(T(x)) e^{-\alpha_{\nu}x} \mathrm{d}x, 
\label{Eq:Fluss}
\end{equation} 
where $\nu$ is the frequency of MIRO's sub-mm or mm channel, $\alpha_{\nu}$ is the length-absorption coefficient, $x$ is the depth measured from 67P's surface, $B_{\nu}$ is the Planck function for blackbody radiation and $T(x)$ is the simulated temperature at depth $x$ (Sect.~\ref{sec:tpm}). It should be noted that the Rayleigh-Jeans approximation is not valid for the temperatures and wavelengths considered in this study \citep{UserManual}. The emission angle is taken into account by replacing the depth $x$ in Eq.~\ref{Eq:Fluss} by the optical path $y$ so that we get
\begin{equation}
I_{\nu}(T) = \int_{0}^{\infty} \alpha_{\nu} B_{\nu}(T(x)) e^{-\alpha_{\nu}y} \mathrm{d}y, 
\label{Eq:Flussy}
\end{equation}
with
\begin{equation}
y = \frac{x}{\mathrm{cos}(\beta)},
\label{Eq:y}
\end{equation}
where $\beta$ is the emission angle. Fig.~\ref{Fig:RadiativeTMEmissionAngle} illustrates the corresponding geometry. Surface reflection is ignored in this simple model, as, with the small refractive index of the cometary material expected from Consert measurements (see Sect.~\ref{Sec:discussion}), a negligible reflectivity smaller than 0.5\% results for a single surface. From the output of Eq.~\ref{Eq:Flussy}, a brightness temperature $T_\mathrm{B}$ is derived by
\begin{equation}
T_\mathrm{B} = \dfrac{h\nu}{k_{B}} \left[\mathrm{ln}\left(1 + \frac{2h\nu^{3}}{I_{\nu}c^{2}}\right)\right]^{-1}.
\label{Eq:T_B}
\end{equation}

\subsubsection{\label{sect:RTM}Ray-tracing approach}
In the case with pebbles, for which the pebble diameter is always comparable to or larger than the wavelength, we cannot neglect reflections at the interfaces between pebbles and void space between the pebbles. Therefore, we model the thermal emission of the comet with a Monte-Carlo ray-tracing approach, applying geometric optics through a three-dimensional periodic structure of specularly reflecting spherical pebbles. The monodisperse spheres are packed randomly at a volume filling factor of $\sim 0.58$, close to a value of 0.55 expected for random loose packing \citep{onoda_liniger_1990}, inside a box with dimensions of 10 pebble radii tangential to the comet's surface and adjusted depths between 30 and 200 pebble radii to keep the total transmission negligible.
    
The complex refractive index of the pebble material is varied in a range $n = 1.0 \ldots 1.3$ for the real part and $k~=~0.001 \ldots 0.01$ at mm wavelength and $k~=~0.005 \ldots 0.1$ at sub-mm wavelength for the imaginary part, which is assumed to be temperature independent.

Rays are traced backwards, from MIRO under the observation angle down to the comet, where they are reflected off the comet's surface or absorbed. A ray's trajectory is changed by refractions or reflections at pebble surfaces as follows. For a ray of given start position $\vec{s}$ and normalised direction $\vec{d}$, first, the closest intersect needs to be found. For every sphere with radius $R$ and centre $\vec{c}$, there is an intersect if
\begin{equation} \label{eq:RaytracingFormulierungBedingRelleWurzel}
	R^2 + \left(\vec{d} \cdot \left(\vec{c} - \vec{s}\right)\right)^2 - \|\vec{s} - \vec{c}\|^2 \geq 0 \text{\space.}
\end{equation}
For any intersect, the distance (or time to travel to the intersect) can be calculated by
\begin{equation}  \label{eq:RaytracingSchnittZeitKugelStrahl}
	t = \, \left(\vec{c} - \vec{s}\right) \cdot \vec{d} - \sqrt{ R^2 - \|\vec{s} - \vec{c}\|^2 + \left(\left(\vec{c} - \vec{s}\right) \cdot \vec{d}\right)^2} \text{\space.}
\end{equation}
The ray interacts with the closest sphere (smallest $t$), where first it is checked whether the ray is reflected off the pebble's surface. Assuming arbitrary polarisation, the reflection probability is given by the average of Fresnel's formulae
\begin{equation} \label{eq:RaytracingFresnel_refl_probability}
	\begin{split}
		p_R &= \frac{R_\mathrm{{perpendicular}} + R_\mathrm{{parallel}}}{2} \\
		&= \frac{1}{2} \left(\frac{n_1 \cos \alpha - n_2 \cos \beta}{n_1 \cos \alpha + n_2 \cos \beta}\right)^{2} + \frac{1}{2} \left(\frac{n_2 \cos \alpha - n_1 \cos \beta}{n_2 \cos \alpha + n_1\cos \beta}\right)^{2}  \text{\space,}
		\end{split}
    \end{equation}
where $\alpha$ and $\beta$ are the angles of incidence and refraction, respectively, which can be calculated from Snell's law in two dimensions. In the case of reflection on the pebble's outside, $n_1 = 1$ and $n_2$ is the pebble's real part of the refractive index. If reflected, the new direction $\vec{d}'$ is calculated by specular reflection (i.e. the angle of incidence equals the angle of reflection) in three dimensions
\begin{equation} \label{eq:RaytracingReflectionLaw}
    \vec{d}' = \vec{d} - 2 (\vec{n} \cdot \vec{d}) \vec{n}  \text{\space,}
\end{equation}
where $\vec{n}$ is the sphere's normal pointing outwards. Else, the ray is refracted according to Snell's law in three dimensions 
\begin{equation} \label{eq:RaytracingSnell3d}
	\vec{d}' = \frac{n_1}{n_2} \vec{d} + \vec{n} \left( - \frac{n_1}{n_2} \vec{d} \cdot \vec{n}  - \sqrt{1 - \frac{n_1^2}{n_2^2} \left(1 - \left(\vec{n} \cdot \vec{d}\right)^2 \right)} \, \right) \text{\space,}
\end{equation}
where $n_1$, $n_2$ and $\vec{d}$, $\vec{d}'$, $\vec{n}$ are as described in Eq.~\ref{eq:RaytracingFresnel_refl_probability} and Eq.~\ref{eq:RaytracingReflectionLaw}, respectively.

Refracted rays can be absorbed inside the pebbles according to the Lambert-Beer law, which provides the transmission probability $p_T$ as a function of the distance $x$ travelled by a ray inside the pebble
\begin{equation}
    p_T(x) = \mathrm{e}^{- \alpha x} = \exp \left( - \frac{4 \pi k}{\lambda} x \right), 
\end{equation}
where $\lambda$ corresponds either to the sub-mm or mm wavelength. If the ray is not absorbed, it can be reflected again, internally, while total internal reflection cannot occur. The probability of reflection is again given by Eq.~\ref{eq:RaytracingFresnel_refl_probability}, but with $n_1$ as the pebble's real part of the index of refraction and $n_2=1$.
    
For every internal reflection, the ray can be absorbed on its path with the same probability. The depth of absorption is always at the point where the ray had its last contact to the pebble's surface. If not absorbed, the ray is finally refracted out of the pebble, according to Eq.~\ref{eq:RaytracingSnell3d}, but with $n_1$ as the pebble's real part of the index of refraction, $n_2=1$ and $\vec{n}$ as the sphere's surface normal pointing inwards.
    
If the ray does not intersect with a sphere, it hits one of the walls of the box containing all spherical pebbles. If this wall is a side wall (its normal is tangential to the comet's surface), a periodic boundary condition is applied. Rays hitting the top of the box are counted as reflected off the comet. Before, they might have been scattered several times by pebbles, i.e. by inner or outer reflection or refraction associated with transmission. If a ray hits the bottom of the box, it is transmitted by the total pebble sample. As the depth of the sample is adjusted such that the number of rays transmitted is negligible, these will be ignored hereafter.

Using the Monte-Carlo method, the ray's starting positions are randomly initialised and decisions for reflection and absorption are made randomly according to their probabilities. Thus, calculations with sufficient numbers of rays converge to reproducible results. To estimate the statistical error of the emission profiles, calculations can be repeated with different random numbers.

For $N$ rays started from the spacecraft under the emission angle, the ray-tracing returns the number of rays, $N_\mathrm{r}$, that were reflected off the comet as well as a list of the depths at which the rays were absorbed. The sample's directional-hemispherical reflectivity is then calculated as the fraction of rays being reflected,
\begin{equation}
    R_\mathrm{dh} = \frac{N_\mathrm{r}}{N},
\end{equation}
and yields the comet's emissivity $1 - R_\mathrm{dh}$, according to Kirchhoff's law \citep{Hapke1993}. The reflectivity ranges from $R_\mathrm{dh} = 0 - 0.035$ at sub-mm to $R_\mathrm{dh} = 0.03 - 0.09$ at mm wavelengths for refractive indices that fit the observations. The rays' absorption depths are binned into layers of one pebble radius thickness, corresponding to the layering of the thermophysical model and result in a normalised emission profile $e(x)$. The comet's total emissivity is thus $(1 - R_\mathrm{dh}) e(x)$.

Reflectivity as well as emission profile $e(x)$ both depend on the refractive index chosen, the emission angle, the pebble radius and the wavelength. For pebbles with a real part of the refractive index of $n = 1.0$ (no scattering), $R_\mathrm{dh}$ is zero and the comets emissivity is unity. Turning on scattering for $n > 1.0$, rays are reflected off the comet, effectively lowering the comet's emissivity and observed brightness temperature.
    
The depth emission profile $e(x)$ from the ray-tracing is multiplied by a modified version of the thermophysical model's intensity profile $B_\nu(T(x))$ (see below), where $B_\nu$ is Planck's function, to get the temperature-dependent emission profile. Integration over depth considering emissivity yields the total comet's intensity at the analysed area. Thus, the synthetic brightness temperature $T_\nu$ is given by
\begin{equation}
    T_\nu = B_\nu^{-1} \left( \left(1-R_{\mathrm{dh}}\right) \int_0^{\infty} B_\nu(T(x)) \, e(x) \mathrm{d}x \right)~.
\end{equation}
Taking into account the additional 3~K cosmic background radiation, with respect to assuming it to be 0~K, changes the brightness temperature of the radiation arriving at MIRO by less than 0.1~K, because of the low surface reflectivity of $\lesssim 1\%$ and the linear superposition of temperatures in the Rayleigh-Jeans limit.

\begin{figure}
    \begin{center}
    	\includegraphics[width =\columnwidth]{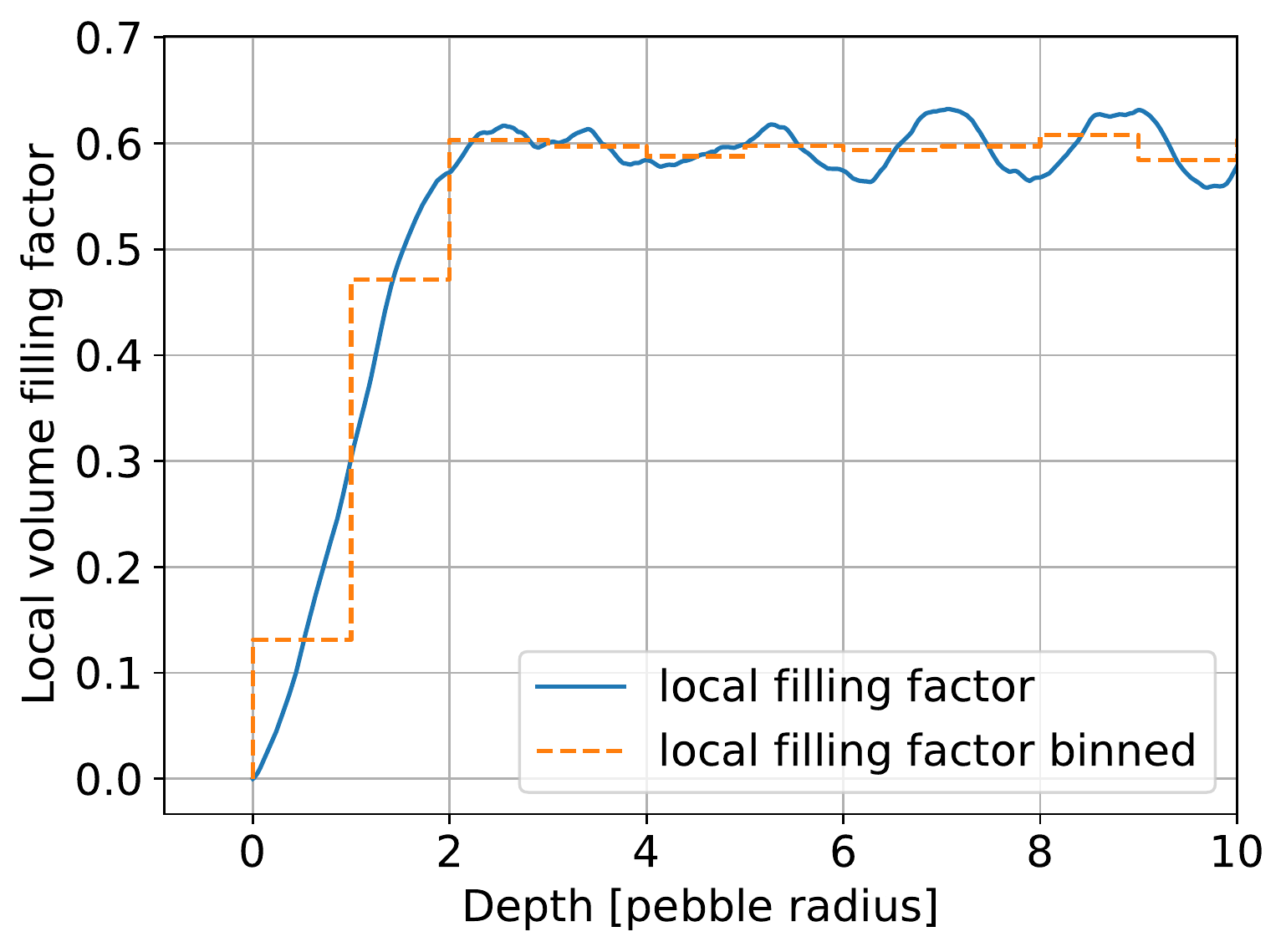}
    		\caption{Local volume filling factor measured by the relative surface area of the circles resulting when cutting horizontally through the pebble structure. Binned graph is averaged over layers of a thickness of one sphere radius.}
    		\label{fig:raytracing_surface_roughness}
    \end{center}
\end{figure}

One correction is made to the thermophysical model's depth-temperature profile (piece-wise constant function of depth), as it is calculated for a homogeneous density, whereas the pebble pile shows a lower local volume filling factor at the surface due to surface roughness (see Fig.~\ref{fig:raytracing_surface_roughness}). This is taken into account by assigning the temperature in the first layer of the thermophysical model to the first and second layer of the pebble sample. Below, the $n$-th thermal layer is assigned to the $(n+1)$-th layer of the pebble sample. Vividly spoken, the derived temperature profile starts at the second layer, while pebble leaking out are assigned the surface's temperature. This is justified, as modified simulations with a reduced volume filling factor in the first two layers (22\% and 79\% of the interior's filling factor, like in Fig.~\ref{fig:raytracing_surface_roughness}) have shown very small differences compared to the modified homogeneous density thermophysical model.

\section{Derivation of the sub-mm/mm optical constants of cometary matter} \label{Sec:OpticalProperties}

\subsection{Length-absorption coefficient for the no-pebble case determined by the Lambert-Beer law} \label{Sec:OpticalPropertiesLambertBeer}

By systematically varying the length-absorption coefficient $\alpha$ in Eq.~\ref{Eq:Flussy} in steps of 0.01 $\mathrm{cm^{-1}}$, a range of synthetic brightness temperatures was derived from the radiative-transfer model described in Sect.~\ref{Sec:lb}. The simulated brightness temperatures were then compared with the ones measured by MIRO to find the $\alpha$ value that best describes the actual brightness temperatures. 
    
\begin{figure}
        \centering
        \includegraphics[width=\columnwidth]{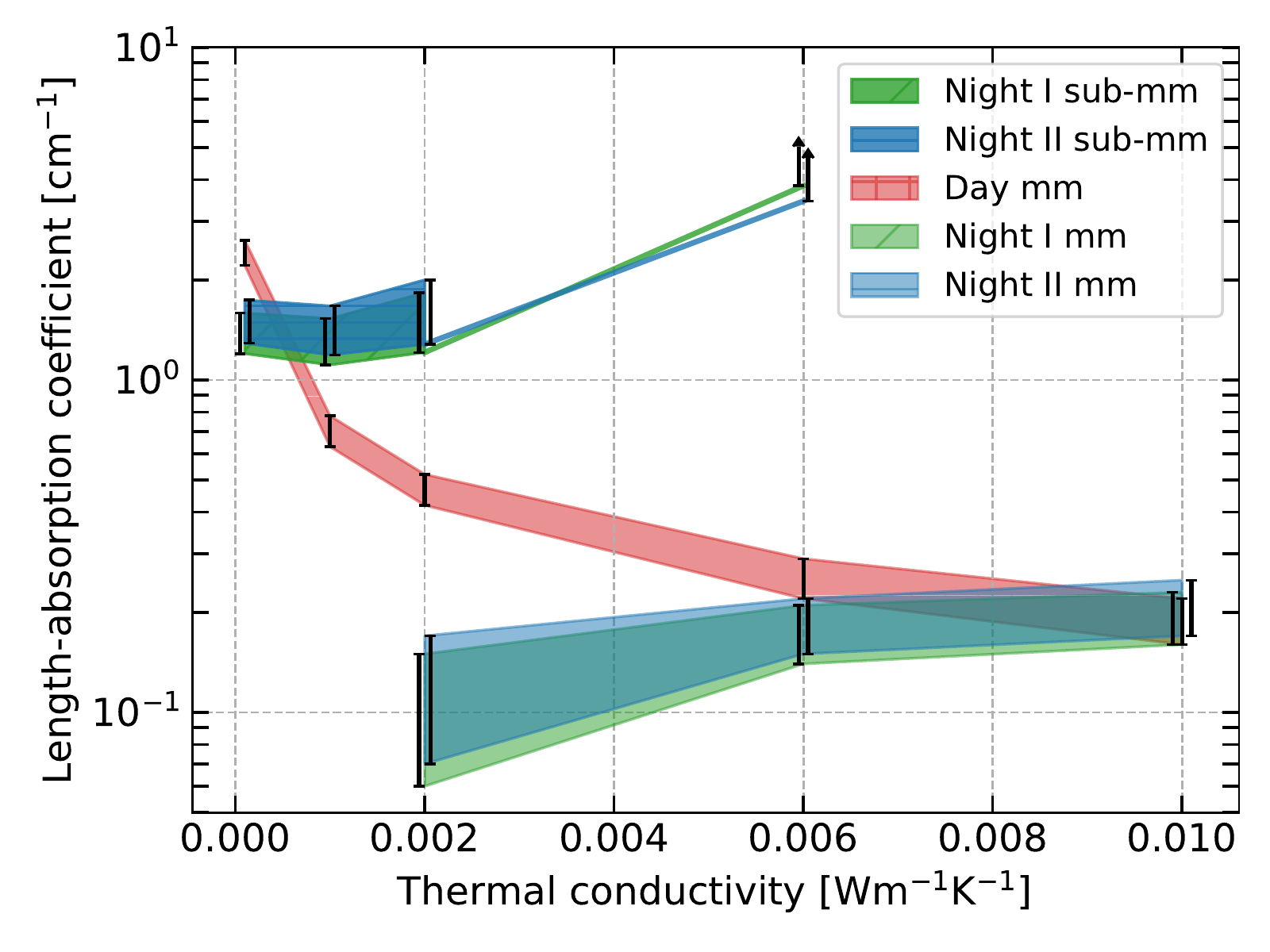}
        \caption{Derived length-absorption coefficients vs. thermal conductivity of the no-pebble material for the sub-mm and mm wavelengths (see Table~\ref{Absorptioncoeff}). We calculated the length-absorption coefficients at the location of the black error bars and connected the data by straight lines for better visibility. In the sub-mm channel, we find only a lower limit for the length-absorption coefficient at $\lambda$ = 0.006~$\mathrm{Wm^{-1}K^{-1}}$ and no solution beyond that thermal conductivity. A solution in both the sub-mm and the mm channel is only found for thermal conductivity values in the range $\lambda$~=~0.002 - 0.006~$\mathrm{Wm^{-1}K^{-1}}$. In this range, a close match between the day and night cases for the mm wavelength is found at a thermal conductivity of $\lambda$~=~0.006~$\mathrm{Wm^{-1}K^{-1}}$ suggesting length-absorption coefficients of $\alpha_{\mathrm{no-pebble, mm}} \approx 0.22~\mathrm{cm^{-1}}$ and $\alpha_{\mathrm{no-pebble, sub-mm}} \geq 3.84~\mathrm{cm^{-1}}$ for the homogeneous dust model.}
        \label{fig:Alpha_vs_Lambda}
\end{figure}

Table~\ref{Absorptioncoeff} and Fig.~\ref{fig:Alpha_vs_Lambda} summarise the results obtained for the length-absorption coefficient in the no-pebble case. All synthetic brightness temperatures as a function of the length-absorption coefficient can be found in Fig.~\ref{Fig:SimulatedTB} in Appendix \ref{AP:LB}. As scattering of the sub-mm/mm waves should be rather unimportant for the no-pebble case, these values are the final results in case of the homogeneous porous dusty matter.

\begin{table*}
	\caption[]{The length-absorption coefficient $\alpha$ in units of $\mathrm{cm^{-1}}$, determined from the Lambert-Beer law as described in Sect.~\ref{Sec:lb}. Presented are the results for different thermal conductivity values $\lambda$ in the no-pebble case. 
	The values for $\alpha_{\mathrm{sub-mm}}$ obtained in the day case are greyed out as they are equal or lower than the values for $\alpha_{\mathrm{mm}}$. (\lowroman{3}) and (\lowroman{4}) mark the solutions we use in the night cases (see Fig. \ref{Fig:SimulatedTemp} in Appendix \ref{AP:LB}). "-" means that there is no length-absorption coefficient found matching the temperatures measured by MIRO. In the no-pebble case, we found a solution in both wavelength channels for thermal conductivity values between $\lambda = 2\cdot10^{-3}$ $\mathrm{Wm^{-1}K^{-1}}$ and $\lambda = 6\cdot10^{-3}$ $\mathrm{Wm^{-1}K^{-1}}$.}
	\centering
	\begin{tabular}{|l|l||c|c|c|c|c|}
		\cline{1-7}
		\multicolumn{2}{|l||}{} & \multicolumn{5}{|c|}{Length-absorption coefficient in $\mathrm{cm^{-1}}$}\\
		\multicolumn{2}{|l||}{} & \multicolumn{5}{|c|} {no-pebble, thermal conductivity in $\mathrm{Wm^{-1}K^{-1}}$}\\
		\multicolumn{2}{|l||}{Case} & $\lambda = 1\cdot10^{-4}$ & $\lambda = 1\cdot10^{-3}$ & $\lambda = 2\cdot10^{-3}$ &$\lambda = 6\cdot10^{-3}$& $\lambda = 1\cdot10^{-2}$\\\hline\hline
		Day & $\alpha_{\mathrm{mm}}$ & 2.21 - 2.63 & 0.63 - 0.78 & 0.42 - 0.52 & 0.22 - 0.29 & 0.16 - 0.22 \\
		& $\alpha_{\mathrm{sub-mm}}$ & \textcolor{gray}{1.88 - 2.62} & \textcolor{gray}{0.52 - 0.78} & \textcolor{gray}{0.33 - 0.52} & \textcolor{gray}{0.17 - 0.29} & \textcolor{gray}{0.12 - 0.22} \\
		\cline{1-7}
        Night \uproman{1} & (\lowroman{3}) $\alpha_{\mathrm{mm}}$ &  - & - & 0.06 - 0.15 & 0.14 - 0.21 & 0.16 - 0.23 \\
		& (\lowroman{4}) $\alpha_{\mathrm{sub-mm}}$ & 1.20 - 1.59 & 1.11 - 1.53 & 1.21 - 1.83 & $\geq$ 3.84 & -\\
		\cline{1-7}
		Night \uproman{2} & (\lowroman{3}) $\alpha_{\mathrm{mm}}$ & - & - & 0.07 - 0.17 & 0.15 - 0.22 & 0.17 - 0.25 \\
		& (\lowroman{4}) $\alpha_{\mathrm{sub-mm}}$ & 1.29 - 1.74 & 1.19 - 1.67 & 1.28 - 2.00 & $\geq$ 3.45 & - \\
		\cline{1-7}
	\end{tabular}
	\label{Absorptioncoeff}
\end{table*}
    
At low thermal conductivity values ($\lambda = 1\cdot10^{-4} - 1\cdot10^{-3}$ $\mathrm{Wm^{-1}K^{-1}}$), sub-surface temperatures are too cold to match the brightness temperatures measured in the mm channel. For high thermal conductivity values ($\lambda = 1\cdot10^{-2}$ $\mathrm{Wm^{-1}K^{-1}}$) the near-surface temperatures are too warm to match the measured sub-mm brightness temperatures.
    
We only found a solution in both wavelength channels for a narrow range of thermal conductivity values between $\lambda = 2\cdot10^{-3}$ $\mathrm{Wm^{-1}K^{-1}}$ and $\lambda = 6\cdot10^{-3}$ $\mathrm{Wm^{-1}K^{-1}}$. 
In this range, we find similar values of the length-absorption coefficients for the day and night cases at mm wavelength only for a thermal conductivity of $\lambda = 6\cdot10^{-3}$ $\mathrm{Wm^{-1}K^{-1}}$. We calculated the final value for the length-absorption coefficient at the sub-mm wavelength by taking the overlap of the two investigated night cases at this thermal conductivity value and get $\alpha_{\mathrm{no-pebble, sub-mm}} \geq 3.84~\mathrm{cm^{-1}}$. For the mm wavelength, we do not find a complete overlap between the two night cases and the day case at $\lambda = 6\cdot10^{-3}$ $\mathrm{Wm^{-1}K^{-1}}$, but the edges of the individual solution ranges touch so that we can estimate $\alpha_{\mathrm{no-pebble, mm}} \approx 0.22~\mathrm{cm^{-1}}$. Thus, the derived length-absorption coefficient for the homogeneous dust model is $\alpha_{\mathrm{no-pebble, sub-mm}} \geq 3.84~\mathrm{cm^{-1}}$ for the sub-mm wavelength and $\alpha_{\mathrm{no-pebble, mm}} \approx 0.22~\mathrm{cm^{-1}}$ for the mm wavelength.

We note that possible daytime shadowing effects in the mm daytime temperatures would result in higher values of the length-absorption coefficient and, thus, increase the difference between the day and night cases.
    
\subsection{Pebble refractive index determined by ray-tracing} \label{sec:OpticalPropertiesRaytracing}

To allow for scattering losses inside the pebble assemblage, we applied the ray-tracing algorithm introduced in Sect.~\ref{sect:RTM} in the pebble case. By systematically varying the complex refractive index $n$ of the pebbles, a range of synthetic brightness temperatures were derived. The simulated brightness temperatures $T_{\mathrm{syn}}$ were then compared with the ones measured by MIRO, $T_{\mathrm{B}}$, to find the refractive indices that best describe the actual brightness temperatures.
    
Fig.~\ref{fig:RT_heatmap_MM} and Fig.~\ref{fig:RT_heatmap_SMM} in the Appendix show the derived temperature differences for all measurement cases, pebble radii, refractive indices and both wavelengths. To find the solution area in the Re($n$)-Im($n$) parameter space, the linearly 2d-interpolated temperature difference $| T_{\mathrm{B}} - T_{\mathrm{syn}} |$ have to be smaller than the temperature error, which is given by
\begin{equation} \label{eq:temp_error_total}
    \Delta T = \sqrt{\Delta T_{\text{B}}^2 + (2\,\sigma_{\text{syn}})^2}
\end{equation}
where $\Delta T_{\mathrm{B}}$ and $\sigma_{\mathrm{syn}}$ are the observational error (see Table~\ref{Tab:TB_MIRO}) and the standard deviation of the synthetic temperature derived by Monte-Carlo ray-tracing, respectively. The latter is calculated by repeating the temperature calculations at a refractive index that best matches the MIRO measurement. The standard deviations differ for each observation case and pebble radius, ranging from $0.099$ to $0.222\,$K. Despite the observational error being considerably larger, this statistical error cannot be neglected, as minor temperature differences play a role in finding solutions at a pebble radius of $3\,$mm.
    
\begin{figure*}
    \centering
    \includegraphics[width=0.75\textwidth]{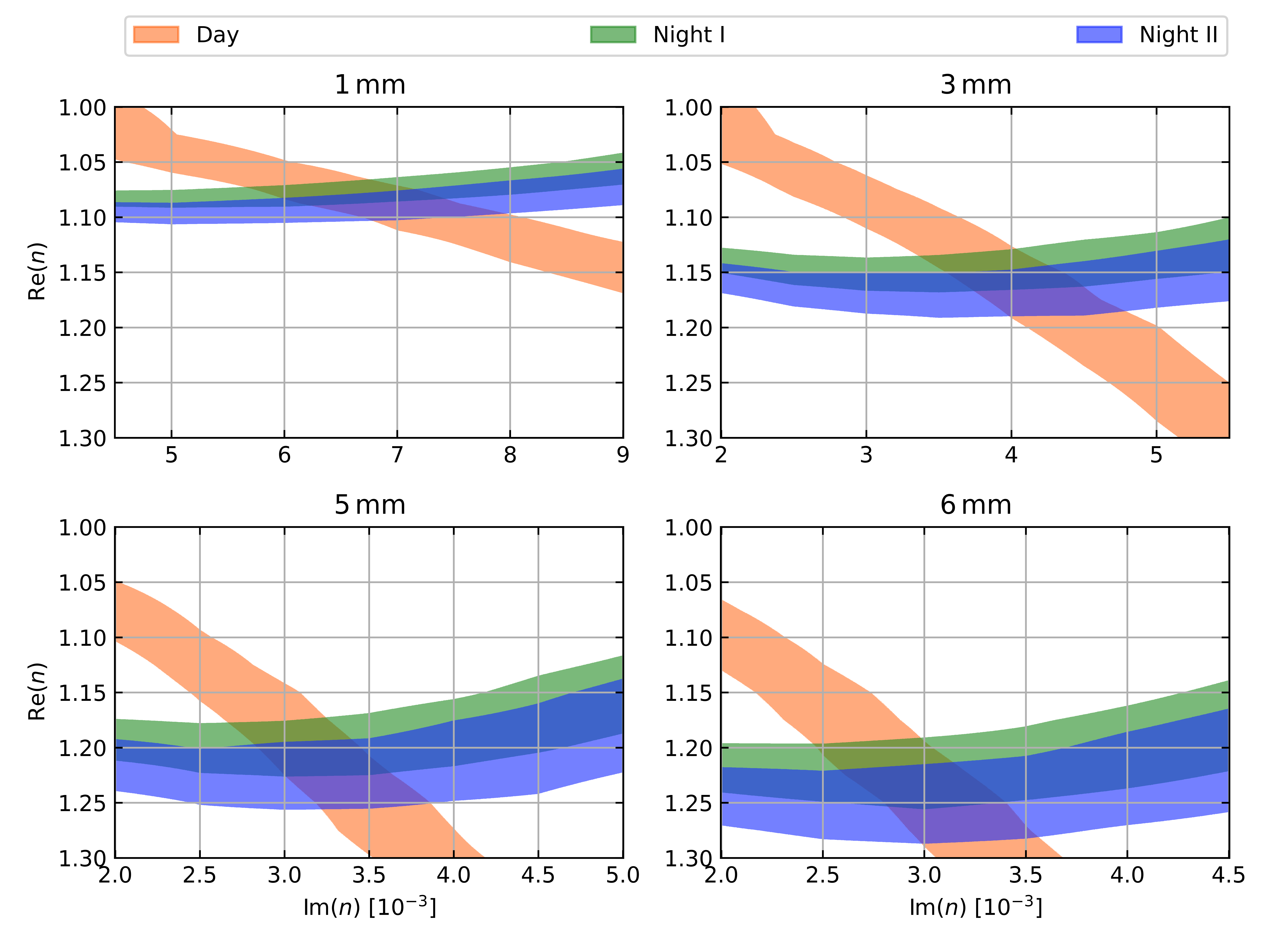}
    \caption{Individual solution areas of the pebbles' refractive indices $n$ for each observation case and pebble radius, calculated by ray-tracing for the mm wavelength observations. The temperature difference values for the grid of discrete refractive indices (see Fig.~\ref{fig:RT_heatmap_MM} in the Appendix) were linearly 2d-interpolated and then compared to the observed MIRO brightness temperature $T_\mathrm{B}$. Areas for which the temperature difference is smaller than the temperature error (Eq.~\ref{eq:temp_error_total}), $ | T_{\text{B}} - T_{\text{syn}} | < \Delta T$, are marked as solution areas.}
    \label{fig:RT_MM_solution_areas}
\end{figure*}

\begin{figure}
    \centering
    \includegraphics[width=0.5\textwidth]{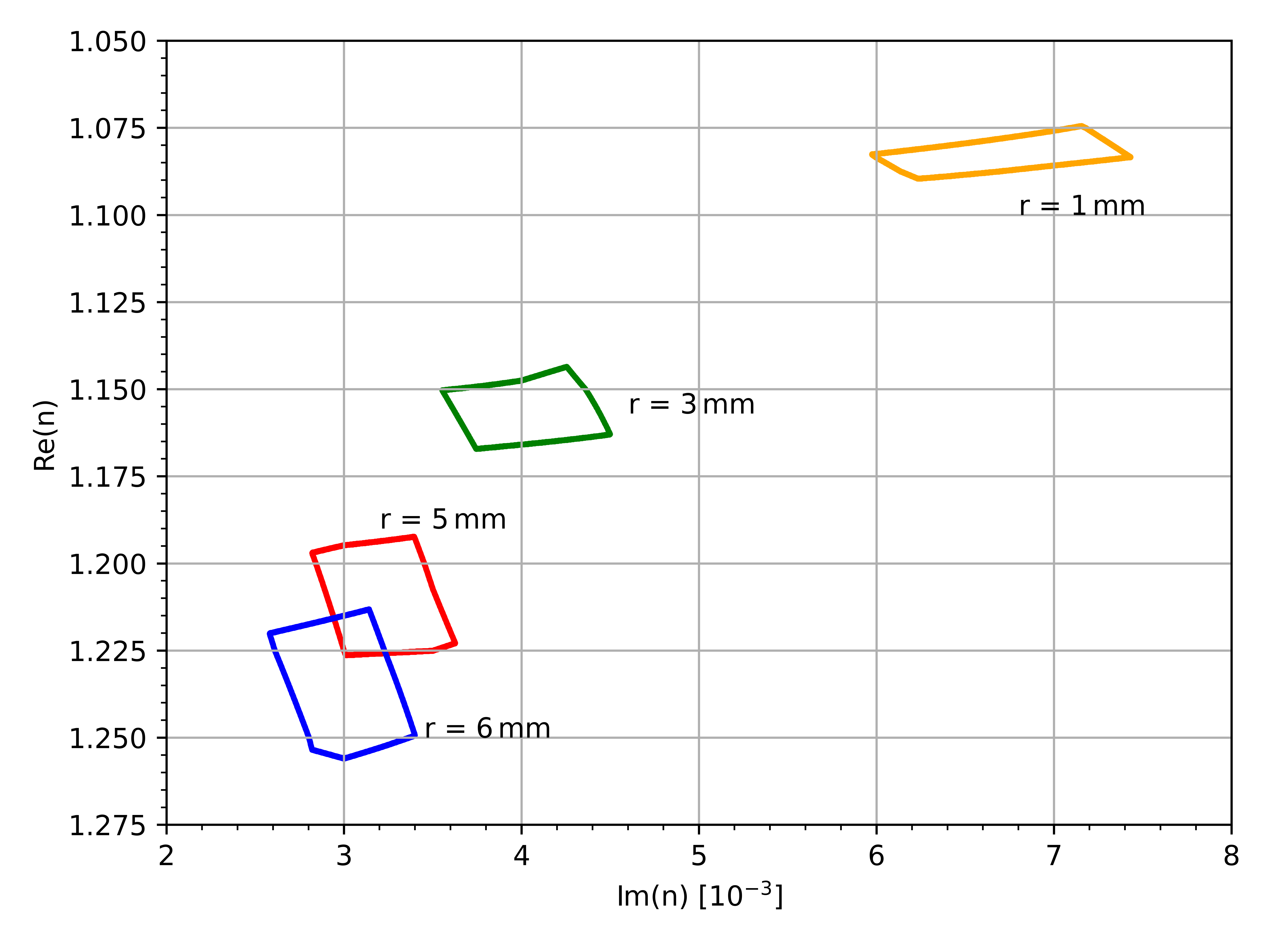}
    \caption{Final solutions of the refractive index of the pebbles with radius $r$ calculated from the overlaps of the individual solution areas in Fig.~\ref{fig:RT_MM_solution_areas}. The optical properties are derived from the comparison to the brightness temperatures measured in MIRO's mm channel.}
    \label{fig:rt_final_solution_areas_n}
\end{figure}

For the mm wavelength, Fig.~\ref{fig:RT_MM_solution_areas} summarises the individual solution areas obtained for all different observation cases and pebble radii. We define the final solution of the pebbles' refractive index as the overlap of all individual solution areas of the three observation cases -- day, night I and night II -- illustrated in Fig.~\ref{fig:rt_final_solution_areas_n} and summarised in Table~\ref{Tab:RealImmarefra}. Final solution areas can be found for all pebble radii and by treating the differences originating from different pebble radii as uncertainties of the refractive index, we find
\begin{equation}
 n_{\mathrm{pebble, mm}} = (1.074 - 1.256) + \mathrm{i} \, (2.580 - 7.431)\cdot 10^{-3} \text{\space.}
\end{equation}
In this range of values, higher real parts and lower imaginary parts of the refractive index correspond to larger pebble radii, as can be seen in Fig. \ref{fig:rt_final_solution_areas_n}. Quantitatively, the refractive index changes from $n = (1.074 - 1.090) + \mathrm{i}\,(5.972 - 7.431)\cdot 10^{-3}$ at $r = 1\,$mm pebble radius to $n = (1.213 - 1.256) + \mathrm{i}\, (2.580 - 3.400)\cdot 10^{-3}$ at $r = 6\,$mm pebble radius (see Table~\ref{Tab:RealImmarefra}).

\begin{table}
\centering
\caption{Limits of the real and imaginary part of the refractive indices of the pebbles, derived from the overlapping areas of Fig.~\ref{fig:RT_MM_solution_areas} for the mm wavelength ($\lambda$ = 1.594~mm). The interdependence of Re($n$) and Im($n$) is shown in Fig.~\ref{fig:rt_final_solution_areas_n}.}
\begin{tabular}{|c|c|c|}
 \hline
  Radius & Real part & Imaginary part ($10^{-3}$)  \\ 
 \hline\hline
 1 mm & 1.074 - 1.090 & 5.972 - 7.431 \\  
 \hline
 3 mm & 1.143 - 1.168 & 3.552 - 4.500\\  
 \hline
 5 mm & 1.192 - 1.226 & 2.819 - 3.627 \\  
 \hline
 6 mm & 1.213 - 1.256 & 2.580 - 3.400 \\  
 \hline
\end{tabular}
\label{Tab:RealImmarefra}
\end{table}
    
We note that, if daytime shadowing would also have a significant influence on the mm emission, and would be included in the thermal modelling, we would find solutions at complex refractive indices that give currently positive deviations in Fig.\,\ref{fig:RT_heatmap_MM} (right panels). This would lead to solutions at larger imaginary parts.
    
In the sub-mm (Fig.~\ref{fig:RT_heatmap_SMM}) wavelength region, we find solutions for the night-time cases only at small pebble sizes and by trend at large refractive indices (both real and imaginary parts). The reason is that close to the surface the temperatures are small enough to account for the small measured brightness temperatures. However, this causes difficulties with our simulations concerning the resolution of the temperature profiles (one pebble radius). Increasing this resolution would require to consider temperature gradients within pebbles. As this is beyond the current complexity of our model, this leads us to discard the sub-mm results from further evaluation. As a result, the solutions in the pebble-model depend on fewer constraints than in the homogeneous model (see Sect.~\ref{Sec:OpticalPropertiesLambertBeer}).

In general, it should be noted that the smallest investigated pebbles with a radius of 1~mm are not much larger than the wavelengths of the MIRO instrument so that the results derived from ray-tracing should be taken with more caution.
We also compared the ray-tracing method to waveoptics simulations in order to check the possible influence of waveoptics effects. As shown in Appendix \ref{AP:WO}, we calculated the absorption losses in a simple layered pebble structure for the mm wavelength channel at normal incidence. We found that for the larger pebbles, the ray-tracing underestimates the average length-absorption coefficient possibly by a factor of up to 1.2-1.3, increasing with increasing real part of the refractive index. For small pebbles and also for the reflected radiation, results depend too strongly on particular waveoptics resonances to deliver definite conclusions. Taking the absorption enhancement into account, would shift the solution areas in Fig.~\ref{fig:rt_final_solution_areas_n} slightly towards smaller imaginary parts of the refractive index. For the sub-mm wavelength, we expect any waveoptics effects to be small because of the strong damping within the pebble material.

\section{Discussion} \label{Sec:discussion}

\subsection{Comparison of the models to the entire diurnal temperature curve}

\begin{figure*}
    \centering
    \includegraphics[width=\textwidth]{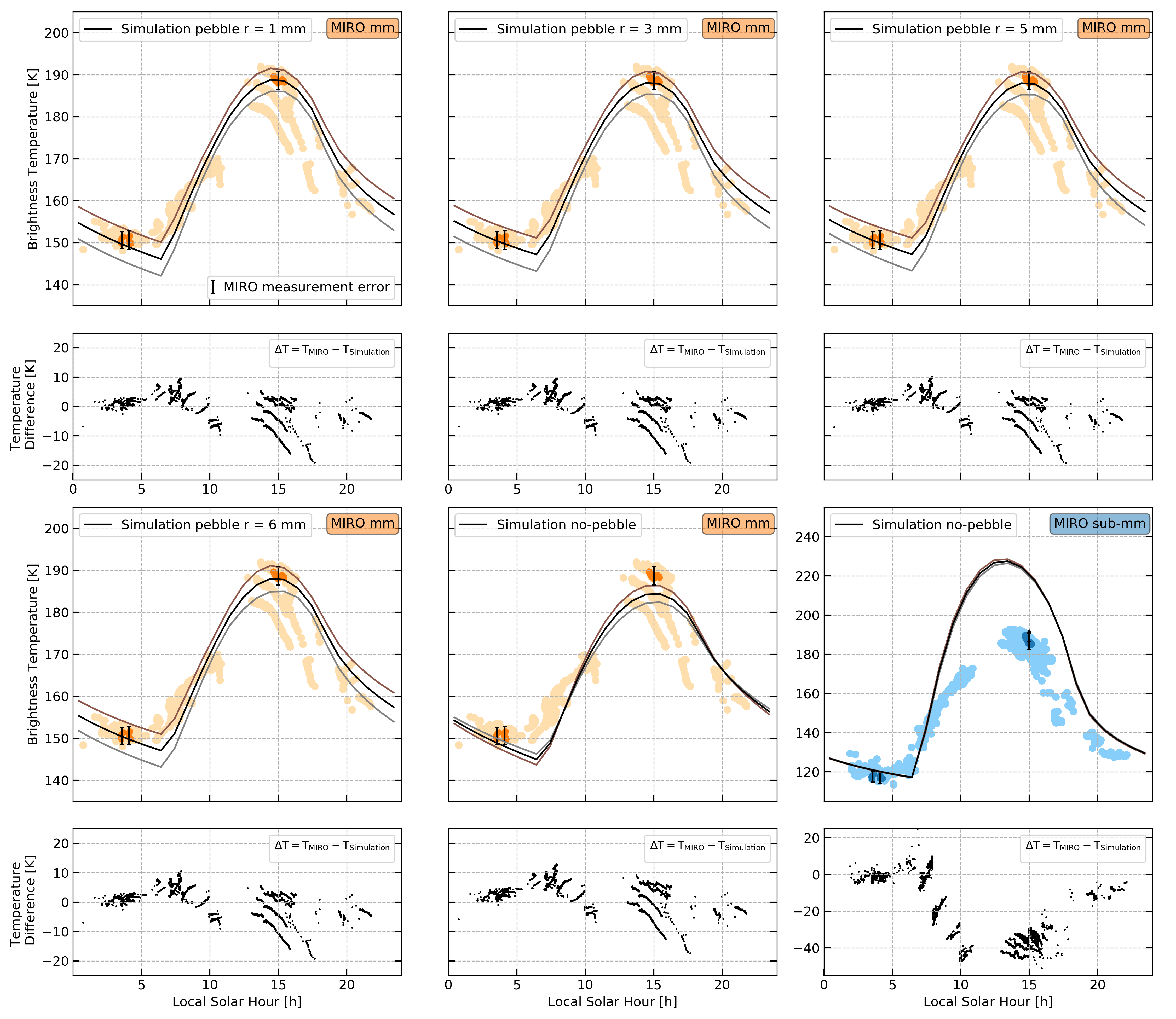}
    \caption{Comparison of the model results to the entire diurnal temperature curve of MIRO, as presented in Fig.~\ref{Fig:Temp_curves_MIRO}. Synthetic brightness temperatures were calculated over the full range of local solar hours, for the range in emission angles (30\textdegree{} to 46\textdegree{}) given by the selected cases (day, night I and II) and for the derived optical properties. In the case of the sub-mm wavelength, only a lower limit for the length absorption coefficient was found and instead of the entire range, only this value is shown here. The selected cases (day, night I and II) are highlighted with darker colours together with the respective measurement error (see Table~\ref{Tab:TB_MIRO}). The upper and lower boundaries of the simulated temperature curves are coloured in brown and grey, respectively, and their mean is illustrated in black. The lower panels show the residuals calculated as the difference between the measurement and the mean value of the simulation. Discrepancies may be due to variations in emission angle and latitude.}
    \label{fig:comparison_models_phase_curve}
\end{figure*}

We calculated the synthetic brightness temperatures over the full range of local solar hours (0 - 24 h), using the derived optical properties and compared them to the entire diurnal temperature curve shown in Fig.~\ref{Fig:Temp_curves_MIRO} and described in Sect.~\ref{Sec:Selection}. This comparison was done for all matching models, the homogeneous and the pebble model, and is illustrated in Fig.~\ref{fig:comparison_models_phase_curve}. The synthetic brightness temperatures were calculated for the range of emission angles (30\textdegree{} to 46\textdegree{}) present for the selected cases (day, night I and II) and for the derived optical properties. 

Our model results agree with the general shape of the diurnal temperature curve for the mm wavelength. The pebble model shows slightly better agreement than the homogeneous model, especially towards the end of the night and at sunrise. Table~\ref{tab:comparison_models_phase_curve} lists the root-mean-square deviations (RMSD) and the mean absolute errors (MAE) of the temperatures derived for the upper and lower boundaries of the simulated temperature curves and their mean value. The mean pebble model possesses smaller RMSD and MAE values than the mean no-pebble model and similar values for the full range of possible optical constants. It should be notetd that the RMSD and MAE values tend to be minimal for pebble radii around 3~mm, which agrees with the pebble size derived by \citet{Blum2017}. In order to estimate the best possible RMSD and MAE values with respect to the noisy MIRO data, we calculated the moving average over the entire diurnal temperature profile and list the corresponding RMSD and MAE values in the last column of Table~\ref{tab:comparison_models_phase_curve}. One can see that RMSD and MAE values of 3.6~K and 2.6~K are the best possible expectation for any model. With our best-achieved model results of 4.5~K and 3.5~K, respectively, we approximate the mean MIRO results with an average accuracy of about 1~K.

\begin{table*}
	\caption[]{Quantitative comparison of the different models to the entire diurnal mm-temperature curve (see Fig.~\ref{fig:comparison_models_phase_curve}). The root-mean-square deviation (RMSD) and the mean absolute error (MAE) were derived for the two boundaries (min and max) of the simulated temperature curves as well as for their mean value. In order to estimate the RMSD and MAE values for a best-fit scenario, a moving average over the entire diurnal phase curve was calculated and is shown in the last column.}
	\centering
	\begin{tabular}{|l||c|c|c|c|c|c|}
		\cline{1-7}
		\multicolumn{1}{|l||}{} & \multicolumn{5}{|c|}{Models} & \\
		\multicolumn{1}{|l||}{} & 1 mm pebble & 3 mm pebble & 5 mm pebble & 6 mm pebble & no-pebble & moving average\\\hline\hline
		\multicolumn{7}{|l||}{RMSD [K]}\\
		max & 5.9 & 5.6 & 5.5 & 5.7	& 5.3 &\\
		mean & 4.7 & 4.5 & 4.6 & 4.7 & 5.2 & 3.6\\
        min	& 5.3 & 5.3 & 5.3 & 5.5 & 5.5 &\\
        \cline{1-7}
        \multicolumn{7}{|l||}{MAE [K]}\\
        max	& 4.5 & 4.2	& 4.1 & 4.3	& 4.4 &\\
        mean & 3.5 & 3.5 & 3.6 & 3.7 & 4.5 & 2.6\\
        min	& 4.4 & 4.4	& 4.5& 4.7 & 4.7 &\\
		\cline{1-7}
	\end{tabular}
	\label{tab:comparison_models_phase_curve}
\end{table*}

The bottom right plot in Fig. \ref{fig:comparison_models_phase_curve} shows a comparison between the sub-mm diurnal MIRO temperature curve and the homogeneous model. The modelled brightness temperatures are much higher during the day than the measured ones. As expected, the diurnal average of the modelled sub-mm temperature is not smaller than that of the mm model, indicating that the measured sub-mm temperatures suffer from daytime shadowing, as discussed in Sect.~\ref{Sec:shadowing}.

In general, discrepancies between modelled and measured brightness temperatures may occur due to deviations in emission angle and latitude. The diurnal temperature curve was filtered for emission angles <60\textdegree{}, i.e. the brightness temperatures could be measured at emission angles from 0\textdegree{} to 60\textdegree{}. Furthermore, the data presented in Fig.~\ref{Fig:Temp_curves_MIRO} are measured in the sub-region 'Imhotep a', which is located between approximately +20\textdegree{} and -40\textdegree{} latitude. The selected and modelled brightness temperatures are located at the equator of 67P.

\subsection{Constraints on composition} \label{Sec:composition}

Using the derived refractive indices for the pebbles (Table~\ref{Tab:RealImmarefra}), we are able to set constraints on the composition of the uppermost layers of the cometary nucleus. The refractive index is a complex number whose real and imaginary parts are mainly linked to the material properties and porosity of the pebbles. The refractive index of cometary analog material is not well known at MIRO frequencies and at low temperatures, but studies have been made on the permittivity (linked to the refractive index) of meteorites (see \citet{HEGGY2012}), which are used as analog to the cometary nucleus refractory dust. For example, \citet{Kofmanaab0639} used the electric properties of carbonaceous and ordinary chondrites to derive constraints on the porosity of the nucleus. Later they expanded the list of potential materials that could match the Consert radar measurements with some organic analog candidates \citep{Herique2016}. All the electric properties used for these studies are in the radar frequency range. Assuming that most geological materials at very low temperature show little variation in the real part of the permittivity with frequency below 1000 GHz, i.e. the lowest-energy optical phonon frequency, we make the assumption that the real part of the cometary analogs considered are the same at radar and microwave frequencies. As the imaginary part has a strong wavelength and temperature dependence even at low temperatures, the same extrapolation cannot be made for it.

A complete study of the constraints on the composition of the nucleus is outside of the scope of this paper, but based on some of the materials presented in Table~7 in \citet{Herique2016}, we can make an educated guess on the volume filling factor of the pebbles and its dependence on the refractory material that they are composed of. We take materials number 7 and 8 in Table~7 of \citet{Herique2016}, namely CR2 carbonaceous chondrite ($\epsilon_r$ = 3.5-4) and carbon with Mg-silicate ($\epsilon_r$ = 2.6-3.1). We choose these as they best fit the radar measurements of the comet, while containing a mixture of chemical components as expected for comet 67P. The sample number 9 in Table~7 of \citet{Herique2016}, namely pure refractory carbon UCAMM-like was not considered here, because the composition of the cometary dust contains some fraction of non-organic material in the form of siliceous minerals as revealed by the COSIMA instrument \citep{Bardyn2017}.

Using Fig.~\ref{fig:rt_final_solution_areas_n}, we are able to determine a range for the real part of the refractive index for the mm wavelength and for pebble radii of 1, 3, 5 and 6 mm. For a pebble radius of 1~mm, the real parts derived are below the values measured by the Consert Radar \citep{Kofmanaab0639}, making it impossible to match the Consert values with these pebbles. We nevertheless plotted the pebble composition for these radii too.

If we assume that the pebbles are made of a three-phase mixture, where one phase is vacuum, one is water ice and the third is the refractory component selected, we can use dielectric mixing laws to evaluate the real part of the permittivity of the mixture as a function of the composition. In order to take into account the error added by the mixing law, we choose to use the Hashin-Shtrikman bounds, which provide an upper and lower limit to the permittivity of the mixture \citep{sihvola1999electromagnetic}. By using these, we are able to represent in a ternary diagram the potential composition of the pebbles for all pebble radii. It is important to know that each refractory candidate used has a range for the real parts of the permittivity. As the real part of the permittivity of the pebbles is quite low, using the highest value would result in an extremely high and unlikely porosity, therefore we will only use the lowest value for each analog. This has the effect of preventing us of providing a lower bound to the volume filling factor and as such we only constrain the upper limit.

\begin{figure*}
	\begin{center}
		\includegraphics[width =0.75\textwidth]{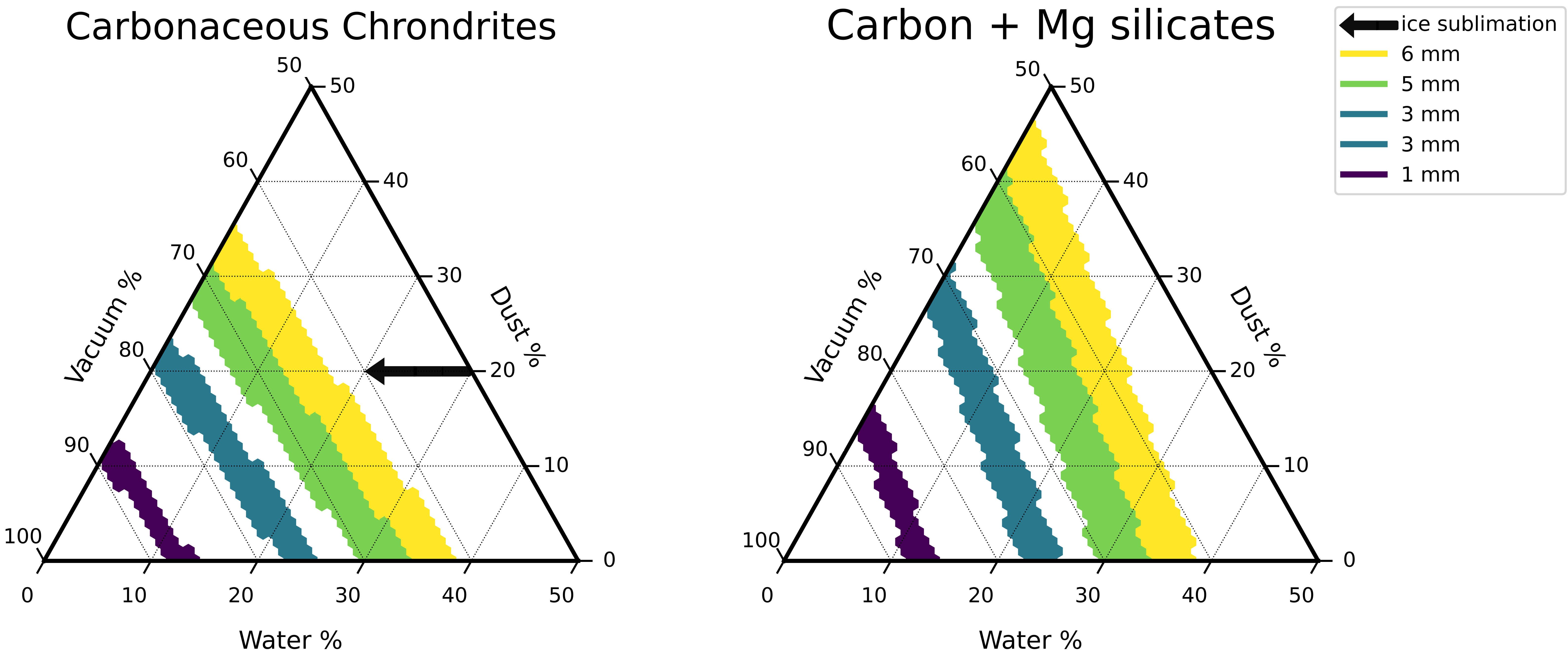}
		\caption{Constraints on the pebble composition by considering the refractory fraction of the comet to be made of carbonaceous chondrites (left) and carbon mixed with magnesium silicates (right). These two compositions were chosen, because they are the most probable ones as constrained by Consert \citep{Herique2016}. The constraints are represented in ternary plots allowing for the representation of the composition of three-phase mixtures. The ternary plots are zoomed in the region with less than 50\% dust and ice, as the pebbles have a high porosity. The constraints obtained are represented by the shaded areas of colour, each representing a given pebble radius. An arrow was also added to represent the sublimation of ice, this sublimation follows the grey lines as the water ice is replaced by vacuum and the dust fraction stays constant.}
		\label{Fig:fillingfactor}
	\end{center}
\end{figure*}

The results are presented in Fig.~\ref{Fig:fillingfactor} for the mm wavelength and the four pebble radii. In general, we see that the carbon mixed with Mg silicates provides the highest volume filling factors, whereas the carbonaceous chrondrite has the lowest, this is due to their respective material permittivities ($\epsilon _r^{\mathrm{CC}}$>$\epsilon _r^{\mathrm{Carbon}}$). We observe that the maximum volume factor increases with increasing pebble radius.

The derived volume filling factors are quite low, at maximum they reach 0.45 for pure carbon mixed with Mg silicates with 6~mm pebbles. Combined with the inter-pebble volume filling factor of 0.58 this results in a total volume filling factor of approximately 0.26, which is very close to the range 0.2-0.25 that was derived for 67P \citep{Patzold2018}. With the 5~mm pebbles, the volume filling factor of the pebbles reaches at most 0.41 resulting in a total volume filling factor of the comet of around 0.24 in the range estimated for 67P. For the 3~mm pebbles, the volume filling factor of the pebbles reaches at most 0.30 resulting in a total volume filling factor of the comet of around 0.17, lower than the range estimated for 67P. Finally for the 1~mm pebbles, the volume filling factor of the pebbles reaches at most 0.16 resulting in a total volume filling factor of the comet of around 0.09 much lower than the range estimated for 67P. The total volume filling factors for the carbonaceous chondrite cases are much lower (the 6~mm pebbles give a total volume filling of 0.20). This leads us to conclude that the smallest pebbles with a radius of 1~mm are unlikely.

Considering that a carbon-dominated material would best correspond to the porosity constraints from the point of view of the real part, we may try to compare the range of derived imaginary parts of the pebble refractive index at $\lambda\sim 1.6$mm (Table~\ref{Tab:RealImmarefra}) to available optical constants of carbonaceous materials. We find that, if extrapolated from $\lambda\sim 0.5$mm, data published by \citet{Jaeger1998} for cellulose pyrolyzed at T$_{pyr}\leq600^\circ$C correspond approximately to the imaginary parts in Table~\ref{Tab:RealImmarefra}. Scaled down with the pebble porosity and additionally reduced by a factor of 2 (Greif et al., in preparation) for the lower temperature at the cometary surface compared to T$_{meas}=300$K, we would obtain, e.g., an imaginary part of about $4\times 10^{-3}$ for the material produced at T$_{pyr}=400^\circ$C. The real part of the refractive index is about 1.8 for this material, which is slightly above that of carbon + Mg silicate, but below that of the CR2 carbonaceous chondrite considered by \citet{Herique2016}.

The data published by \citet{Zubko1996} for condensation and combustion products \citep{Colangeli1995}, in contrast, have imaginary parts that are two orders of magnitude too high, given that their temperature dependence at T$\geq$160K is quite weak \citep{Mennella1998}. A material with a higher content of organics is represented by optical constants published by \citet{Pollack1994}, based on measurements on tholins by \citet{Khare1984}. These data are characterised by a steep decay of the imaginary part from the sub-millimetre to the millimetre range, which might better correspond to the large difference in brightness temperatures (optical depths) measured by MIRO for the two wavelengths than the \citet{Jaeger1998} data. However, the real part of the refractive index given in these data ($Re(n)\sim2.2$ for $\lambda>300\mu m$) would not be compatible with the Consert measurements at realistic porosities.

The same discussion applies more or less to the length-absorption coefficients found in the homogeneous-surface case. If translated into an imaginary part of the refractive index for the $\mu$m-sized dust, this value is larger than for the pebble material by a factor of 3-4 at the millimetre wavelength, but still in the range of less graphitic or organic carbon. The lower limit at the sub-millimetre wavelength indicates again a rather steep rise of the extinction constant towards shorter wavelengths. If this would apply, it could point to a spectral characteristics similar to that of the tholins mentioned above.

\subsection{Constraints on formation and structure}
Starting with the two major planetesimal-formation scenarios (see Sect. \ref{sec:formation}), we idealised the cometary sub-surface region as either consisting of macroscopic dust pebbles, which themselves are made up of microscopic dust grains, or entirely of homogeneously-packed microscopic particles. This distinction leads to differences in the thermal conductivity (see Eqs. \ref{eq:pebblehc} and \ref{eq:nopebblehc}) and to a distinct heat transfer during the diurnal solar cycle and, thus, to measurable surface and sub-surface temperatures \citep[][]{Bischoff2021}.

In the case of the homogeneous sub-surface material with no pebbles, we found only one solution for the length-absorption coefficient at a thermal conductivity of $\lambda = 6\cdot10^{-3}$~$\mathrm{Wm^{-1}K^{-1}}$, which simultaneously explains the sub-mm and mm observations.

As was shown by \citet{Bischoff2021}, the diurnal temperature variation, particularly at or close to the surface, is not a good measure of the thermal conductivity of the sub-surface material. In their Figure 3, \citet{Bischoff2021} compare the surface temperatures of a pebble structure with those of a homogeneous sub-surface composition and find a close match within a few K between the two. They propose to use the dependency of the temperature at sunrise on the insolation at noon as a proxy for the underlying heat-conductivity process. A strongly temperature-dependent heat-conductivity, as in the case of large pebbles, results in a steep increase of morning temperatures with increasing insolations at noon, whereas there is almost no such dependency for a temperature-independent heat-conduction process. Unfortunately, neither the MIRO data nor those of the VIRTIS instrument onboard Rosetta allow such a comparison.

In summary, the straightforward explanation of the MIRO data by a pebble model with pebble radii in the millimetre range, the close match of the thermal conductivity in the no-pebble model with the radiative conductivity for a temperature of $\sim 150$~K and pebble radii of $5~\mathrm{mm}$ \citep[see Figure 1 in][]{Bischoff2021} and the fact that such an agreement was predicted by \citet{Bischoff2021} leads us to the conclusion that the sub-surface of comet 67P most likely consists of pebbles.

We applied our thermophysical model for a hierarchical makeup of the cometary sub-surface with pebble radii ranging from 1~mm to 6~mm. We found formal solutions for the optical constants for all pebble radii in this range. However, comparing the derived real parts of the refractive index with measurements of the Consert instrument and the derived total volume filling factors of the comet with previous estimates, we conclude that the smallest pebbles with a radius of 1~mm are unlikely to constitute the cometary matter. A size range of pebble radii between 3~mm and 6~mm was already deduced by \citet{Blum2017}. It should, however, be noted that the upper limit of the pebble radii is less well-defined, because for large pebbles the underlying model assumption of isothermality breaks down.

In conclusion, our results point strongly towards a pebble makeup of the cometary sub-surface, with pebble radii of a few millimetres. This makeup can readily be explained by a formation of comet 67P (or its parent body) through a gently gravitational collapse of a pebble cloud \citep[][]{JohansenEtAl2007}, which currently is the favoured model of planetesimal formation.

\section{Results in the context of protoplanetary and debris disc observations} \label{Sec:application}
The derived physical properties of the dust/pebbles in the upper layers of comet 67P provide fundamental constraints for the analysis of protoplanetary and debris discs in two respects.
First and foremost, the interaction of electromagnetic radiation with the dust phase is determined by their wavelength-dependent optical properties, sizes distribution and internal structure. Thus, any quantitative analysis of observations of the continuum radiation of these discs in the optical to thermal reemission regime relies on the assumptions about these dust properties, including as basic property the mass of the dust present in discs \citep[e.g.,][]{2005ApJ...631.1134A}.
Second, dust plays a key role in various physical processes that govern the evolution of protoplanetary discs to planetary systems. Here, the absorption and reemission efficiencies of the dust particles determine the temperature and, thus, the hydrostatic structure of these discs, while primarily the mass and size distribution as well as the structure of the grains allow describing the dynamics of the gas-dust interaction and, thus, the processes that eventually lead to the formation of planets \citep[e.g.,][]{2020MNRAS.493.1788G}.

Despite the steady progress of theoretical, observational and laboratory studies, 
there still exist major uncertainties about the above dust properties in protoplanetary and debris discs.
For example, by analyzing the sub-mm/mm slope of the Rayleigh-Jeans tail of the continuum spectral energy distribution 
the basic process of grain coagulation has been verified in many observational protoplanetary disc studies since the early 1990's \citep[e.g.,][]{1991ApJ...381..250B,2014prpl.conf..339T}.
Moreover, the theoretically predicted segregation of grains as a function of their size in radial and vertical distribution has been confirmed 
\citep[e.g.,][]{2013A&A...553A..69G,2015ApJ...813...41P,2016A&A...588A..53T,2020A&A...642A.164V}. However, while these studies often revealed grains of at least millimeter size, the observed degree of sub-mm/mm polarisation of scattered thermal reemission radiation favour a maximum grain size of $\approx$ 0.1\,mm \citep[e.g.,][]{2018ApJ...865L..12B,2018ApJ...860...82H,2020MNRAS.496..169L,2020ApJ...900...81O}. A certain degree of porosity of the grains defuses the problem concerning the degree of linear polarisation \citep[e.g.][]{2019ApJ...885...52T,2019A&A...627L..10B,2020A&A...640A.122B}. Nevertheless, a consistent solution that also takes the observed sub-mm/mm spectral index into account is pending \citep[see, e.g.,][]{2021A&A...648A..87B,2021ApJ...913..117U}.

Within the above context of protoplanetary and debris discs, the results obtained with MIRO are of particular interest in various respects. As they characterise pristine material with characteristics imprinted by processes during the planet formation phase, the subsequent evolution potentially altered the material, providing constraints on the overarching question about the yet weakly constrained connection between the protoplanetary and debris disc phase \citep[e.g.,][]{2021arXiv211106406N}. First, while the optical properties in the sub-mm/mm regime are usually determined on the basis of chemical compositions derived from characteristic emission and absorption features in the optical to infrared wavelength range (below $\lesssim$ 100\,\micron), the MIRO observations provide direct, stringent constraints. This is particularly important, because the estimate of the dust mass in protoplanetary and debris discs and, thus, questions regarding related planet formation scenarios in the former and the collisional evolution in the latter case sensitively depend on this fundamental parameter. Comparing the derived optical properties to those of currently assumed dust compositions in protoplanetary discs \citep[e.g., for the analysis of the ALMA DSHARP survey;][]{2018ApJ...869L..41A}, we find similar values of the imaginary part of the refractive index. This is interesting, because our analysis relies on a rather high porosity of about 60\%, while lower values are assumed in young discs \citep[e.g.,][and references therein]{2018ApJ...869L..45B} which is in agreement with constraints from sub-mm/mm polarisation observations \citep[][]{2021A&A...648A..87B}. Taking the volatile nature of water ice and its depletion due to photosputtering during the debris disc stage into account, the MIRO results might thus allow constraining the relative abundance of the remaining species possessing different millimeter emissivities.

Second, the derived grain size distribution shows an apparent overabundance of grains with 3-6\,mm radius, while the observations even exclude $\approx$1\,mm grains. This finding is in agreement with the scenario in which planetesimals form via gravitational collapse of small pebbles in vortices \citep[e.g.,][]{2021ApJ...913...92R}, pressure bumps \citep[e.g.,][]{2020A&A...635A.105P}, or clusters of aeoredynamically coupled solids initiated by streaming instabilities \citep[e.g.,][]{2017ApJ...847L..12S}. At the same time, ALMA high angular resolution continuum images of protoplanetary discs seen edge-on support this scenario \citep[][]{2020A&A...642A.164V}. The authors find that the studied discs are already partly optically thick in the sub-mm/mm range, implying significant scattering by grains that have grown to mm/cm sizes \citep[see also, e.g.,][]{2008ApJ...674L.101W}.
Furthermore, these observations are in disagreement with pure radial drift models, but require mechanisms such as the predicted pressure traps to decelerate or even stop the radial drift.
Similarly, several rings found in protoplanetary discs within the ALMA DSHARP survey do not only show strong evidence for particle trapping, but allow excluding grains $\gg$1\,mm if low turbulent velocities are assumed \citep[][]{2018ApJ...869L..46D}.
On the other hand, \citet{2018ApJ...869L..45B} derive a minimum grain size of 2\,mm
for the particles in the rings in the disc of HD~163296 from the same survey. Eventually, the overabundance of particles of selected sizes -- if compared to a grain size distribution resulting from an ideal collisional cascade -- may become observable as a result of collision events in debris discs. Here, one would expect a change of the slope of the spectral energy distribution in the wavelength region in which grains of that specific size emit most efficiently \citep[for a potential observation of this phenomenon see][]{2020EPJWC.22800015L}.

\section{Conclusion} \label{Sec:Conclusion}
In this work, we fitted the sub-mm/mm MIRO measurements of the thermal emission of comet 67P's sub-surface with synthetic brightness temperatures derived from a thermophysical model and radiative-transfer models. By assuming the thermophysical model, we were able to determine optical properties of the near-surface material of comet 67P. 

According to the two major formation scenarios for planetesimals, we simulated a dust layer composed of pebbles and a homogeneous dust layer composed of smaller grains. In the latter case, we applied a radiative-transfer model based on the Lambert-Beer law and only found a solution for the optical properties at a thermal conductivity of $\lambda = 0.006$~$\mathrm{Wm^{-1}K^{-1}}$, suggesting length-absorption coefficients for the homogeneous dust model of $\alpha_{\mathrm{no-pebble, sub-mm}} \geq 3.84~\mathrm{cm^{-1}}$ for the sub-mm wavelength and $\alpha_{\mathrm{no-pebble, mm}} \approx 0.22~\mathrm{cm^{-1}}$ for the mm wavelength.

For the pebble case, we applied a ray-tracing algorithm, because the pebble size is larger than the two MIRO wavelengths and scattering effects cannot be neglected. In the ray-tracing simulations, we varied the complex refractive index of the pebbles and derived values of $n_{\mathrm{pebble, mm}} = (1.074 - 1.256) + \mathrm{i} \, (2.580 - 7.431)\cdot 10^{-3}$ for the mm wavelength and pebble radii between 1~mm and 6~mm. The sub-mm results had to be discarded, because the thermal emission measured in this short wavelength originates from the very shallow sub-surface of the comet and modelling temperature gradients within one pebble is beyond the current complexity of the thermophysical model.

The obtained optical properties were used to derive synthetic brightness temperatures over the full range of local solar hours, which were then compared to the entire diurnal MIRO temperature curve. The comparison shows a good agreement between the models and the general shape of the MIRO diurnal temperature curve for the mm wavelength and illustrates the problem of day-time shadowing for the sub-mm wavelength. Although the wavelength difference between the sub-mm and mm channels is only a factor of 3, shadowing might play a much less important role in the mm data, because the penetration depth derived from the calculated length absorption coefficients is about $4.5~\mathrm{cm}$, which corresponds to several pebble radii. In contrast, the penetration depth for the sub-mm wavelength is less than about $0.3~\mathrm{cm}$, which is on the order of the pebble radius. If the roughness is of the order of the pebble size as well, the sub-mm data will be affected by shadowing effects, while the mm data will hardly be affected.

The obtained real part of the refractive index was used to derive constraints on the composition and volume filling factor of the pebbles. Pebbles are assumed to be made of a three-phase mixture consisting of vacuum, water ice and a refractory component, where the latter is represented by carboneaceous chondrites as well as carbon mixed with magnesium silicates taken from \citet{Herique2016}. We find that carbon mixed with magnesium silicate provides the highest inner-pebble volume filling factors and by taking into account the inter-pebble volume filling factor, we found a good agreement with the total volume filling factor of comet 67P for pebble sizes of 5~mm and 6~mm. This comparison with the derived total volume filling factors of the comet and results from the Consert instrument leads us to conclude that the smallest pebbles with a radius of 1~mm are unlikely to constitute the cometary matter. Thus, our findings support the range of pebble sizes suggested in \citet{Blum2017}.

The discussion of our results for the homogeneous sub-surface material with no pebbles and the pebble model points towards a pebble makeup of the cometary sub-surface as e.g., the resulting thermal conductivity for the no-pebble case matches the radiative conductivity for 5~mm pebbles in the investigated temperature range. In general, the underlying heat-conductivity process and thus the sub-surface structure can only be inferred from diurnal surface temperature measurements, when the dependency of the sunrise temperature on the insolation at noon is measured \citep{Bischoff2021}. Such measurements are, unfortunately, not available for comet 67P.

The derived optical and physical properties of the sub-surface of comet 67P provide valuable information in the context of protoplanetary-disc and debris-disc observations as e.g., the estimation of dust masses in these discs depends on the optical properties.

\section*{Acknowledgements}
Data sets of the MIRO instrument have been downloaded from the ESA Planetary Science Archive. The authors acknowledge the Principal Investigator Mark Hofstadter of the MIRO instrument on-board the Rosetta mission for providing data sets in the archive. 

This work was funded by the Deutsche Forschungsgemeinschaft (DFG, German Research Foundation) in the framework of the research unit "Debris Disks in Planetary Systems" (grant 278211407, FOR2285), sub-grants BL 298/24-2 (JBl), MU 1164/9-2 (HM) and WO 857/15-2 (SB), in the framework of the D-A-CH program (BL 298/26-1, GU 1620/3-1; JBl, BG, AL), by grant BL 298/27-1 (JBl and DB) and by grant HO 5868/1-1 (HM and SH). JBü and JBl thank DLR -- German Space Agency for support under grant 50WM1846.  

\section*{Data Availability}
The data underlying this article will be shared on reasonable request to the corresponding author.




\bibliographystyle{mnras}
\bibliography{Bibliographie}




\appendix

\section{\label{AP:TP}Temperature Profiles}
\begin{figure}
    		\includegraphics[width=\columnwidth]{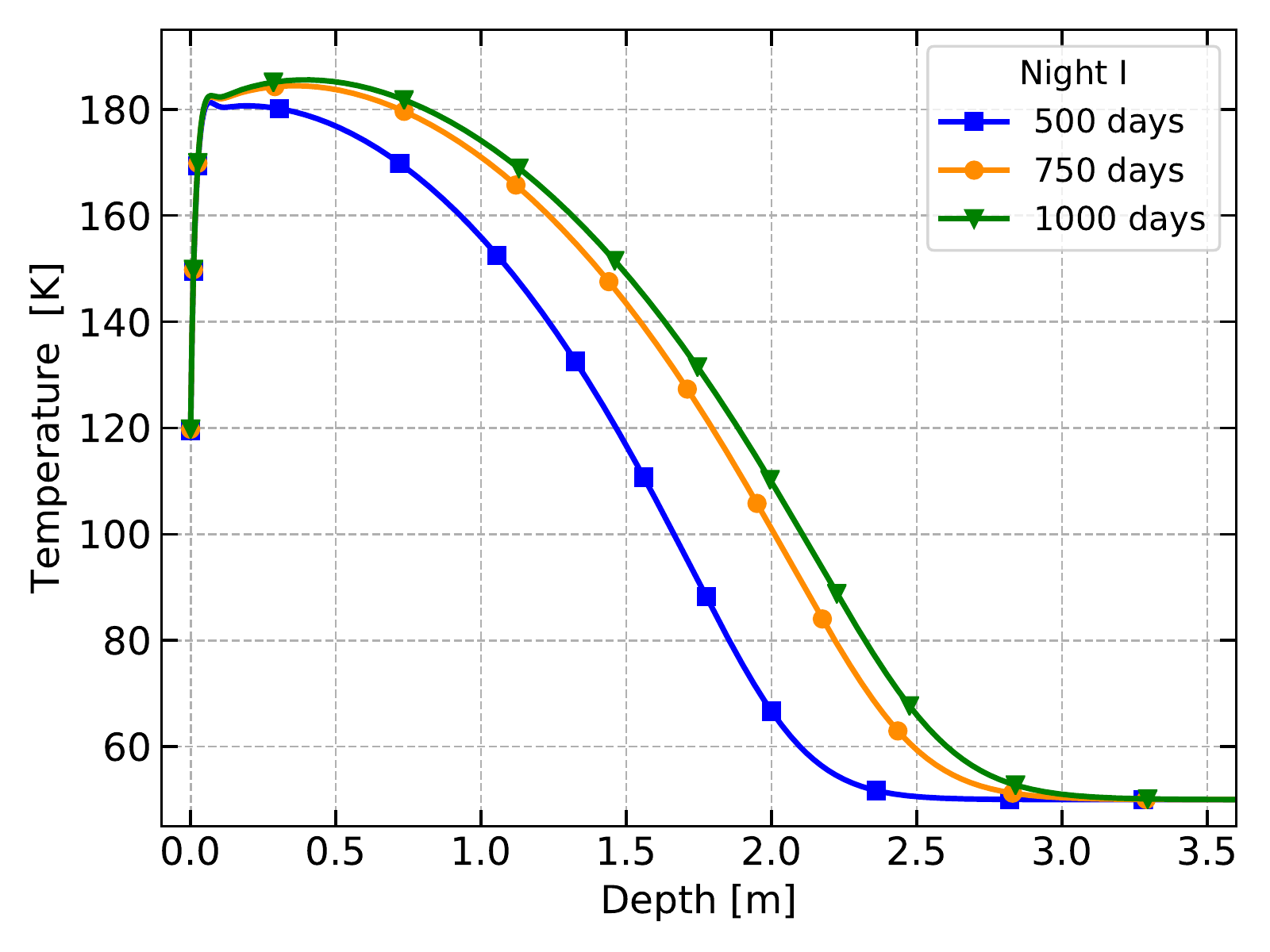}
    		\includegraphics[width=\columnwidth]{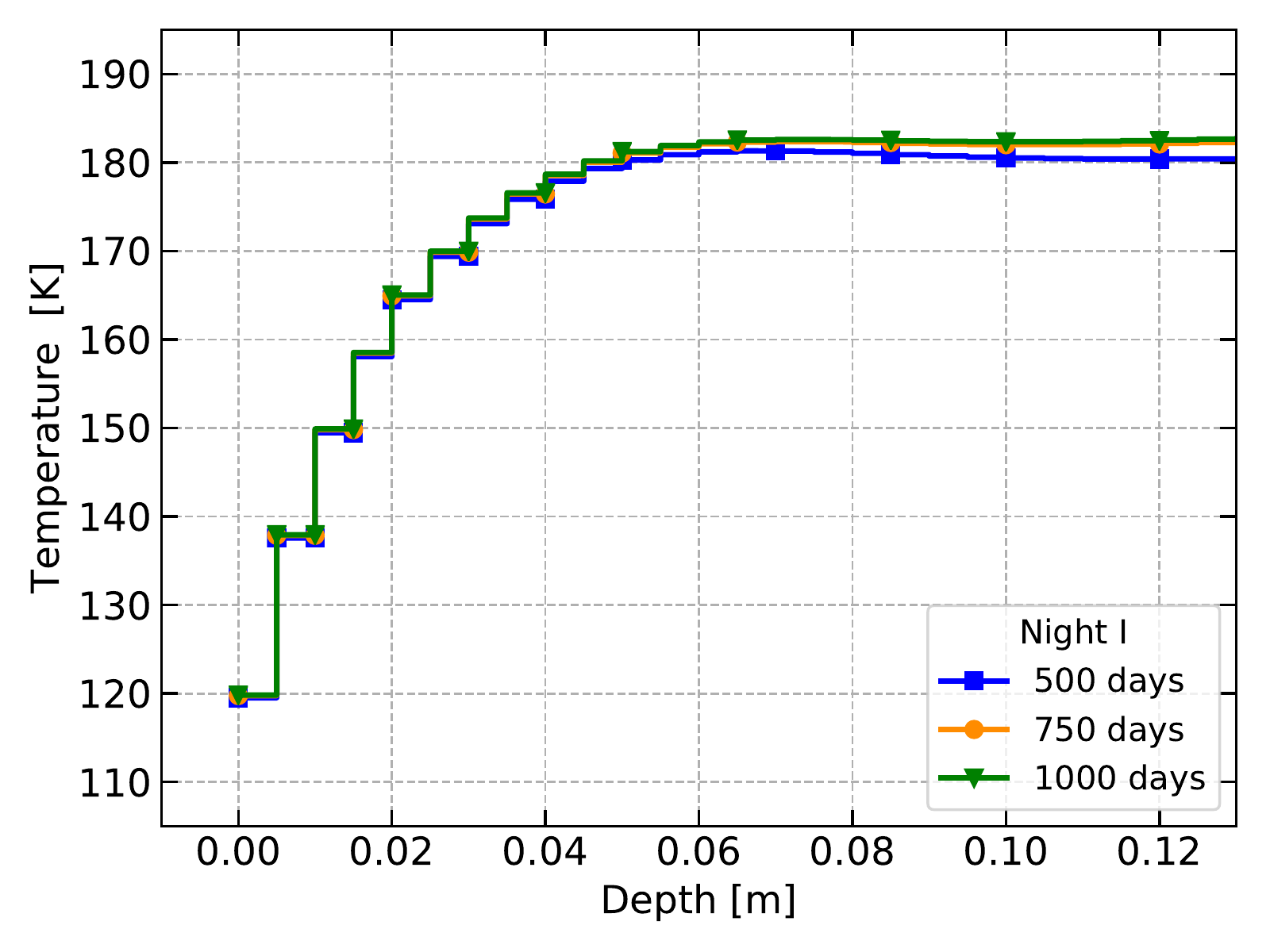}
	\caption[]{Derived temperature profiles for 500, 750 and 1,000 cometary days (see Fig. \ref{Fig:SimulationNumberofCometaryDays}) for the night \uproman{1} pebble case with $r~=~5$~mm. The upper panel shows the whole temperature-depth profile, whereas the lower panel zooms into those sub-surface regions MIRO is approximately sensitive to. For longer simulation times, the heatwave propagates deeper into the comet's interior. However, the near-surface temperatures are quite similar for the 750 and 1,000 cometary days simulations, while the 500 cometary days run shows slightly lower temperatures, due to the initial conditions.}
	\label{Fig:Temperatureprofiles_Night1}
\end{figure}

\begin{figure*}
	\begin{center}
		\includegraphics[width=0.8\textwidth]{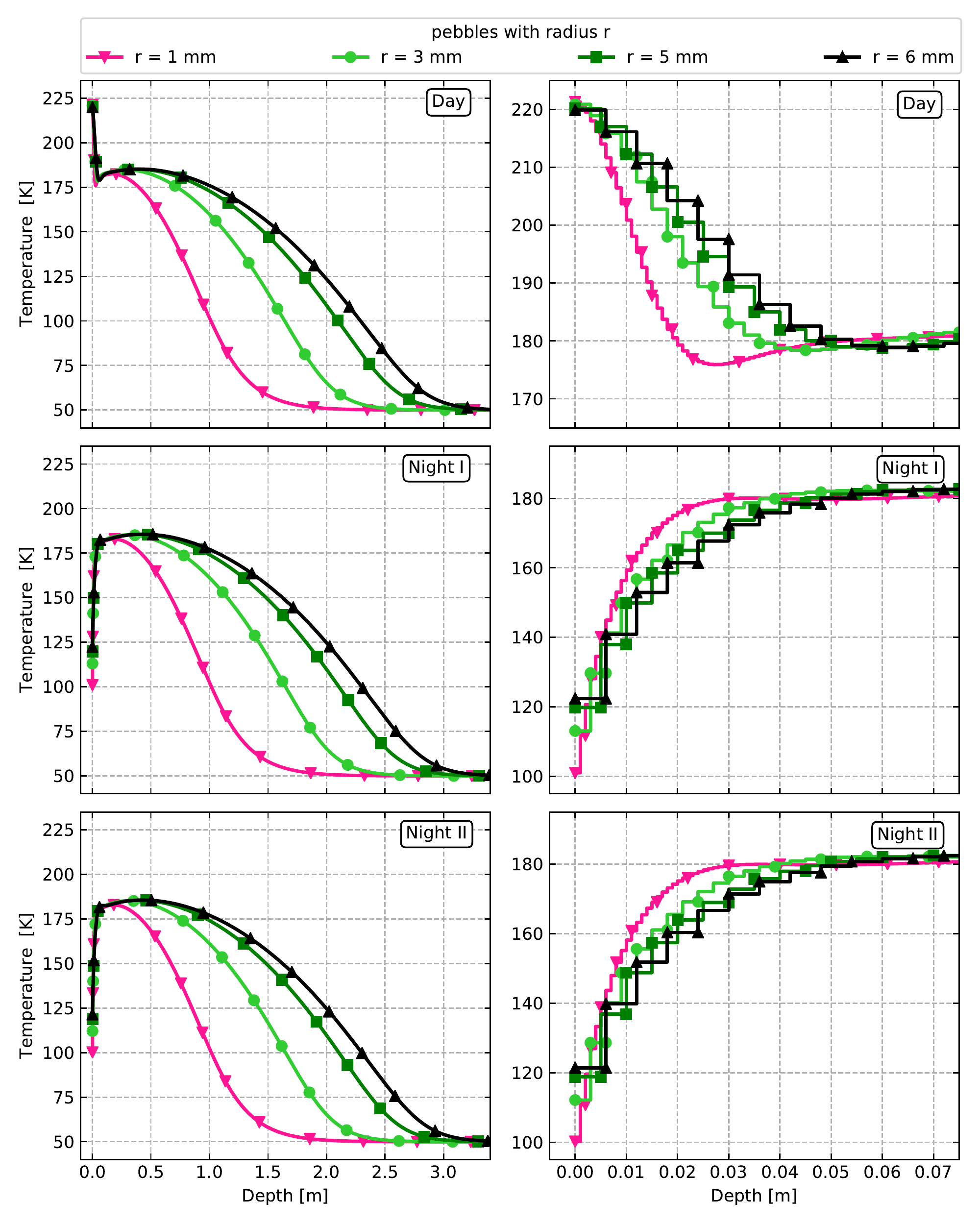}
		\caption{Overview of the temperature profiles derived for the pebble case with different pebble radii $r$.}
		\label{Fig:TemperatureProfiles_pebble}
	\end{center}
\end{figure*}

\begin{figure*}
	\begin{center}
		\includegraphics[width=0.8\textwidth]{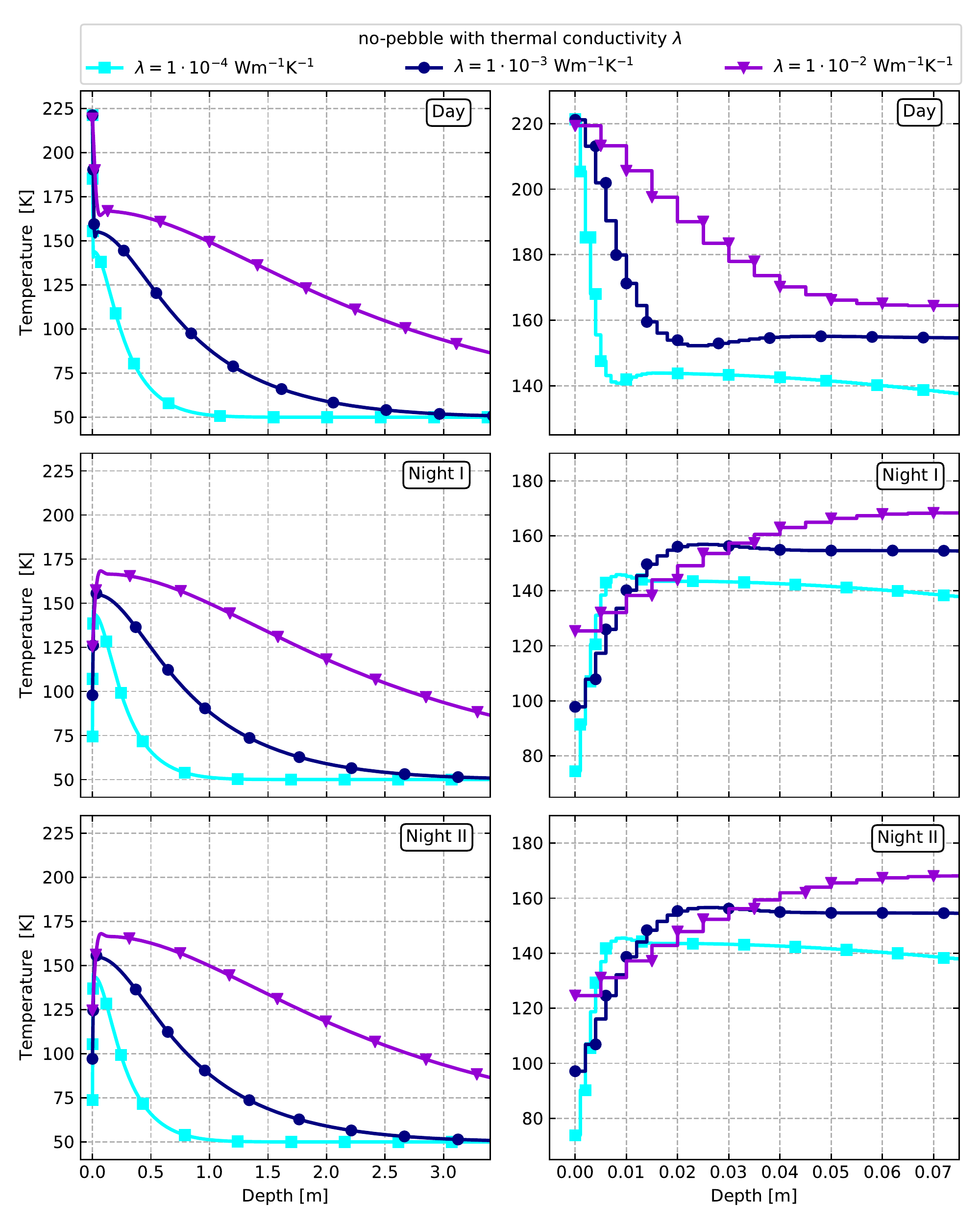}
		\caption{Overview of the temperature profiles derived for the no-pebble case with different values for the thermal conductivity $\lambda$.}
		\label{Fig:TemperatureProfiles_nopebble}
	\end{center}
\end{figure*}

Fig.~\ref{Fig:Temperatureprofiles_Night1} shows the temperature-depth distribution for the night \uproman{1} case, pebbles with 5~mm radius and three different simulation durations of 500, 750, and 1,000 days, respectively. Figs.~\ref{Fig:TemperatureProfiles_pebble} and \ref{Fig:TemperatureProfiles_nopebble} show a comparison of all simulations with pebbles (Fig.~\ref{Fig:TemperatureProfiles_pebble}) and in the no-pebble case (Fig.~\ref{Fig:TemperatureProfiles_nopebble}). 

\section{\label{AP:SBT} Synthetic brightness temperatures and resulting optical constants}

\subsection{\label{AP:LB} Lambert-Beer approach}

\begin{figure}
	\begin{center}
	\includegraphics[width=1\columnwidth]{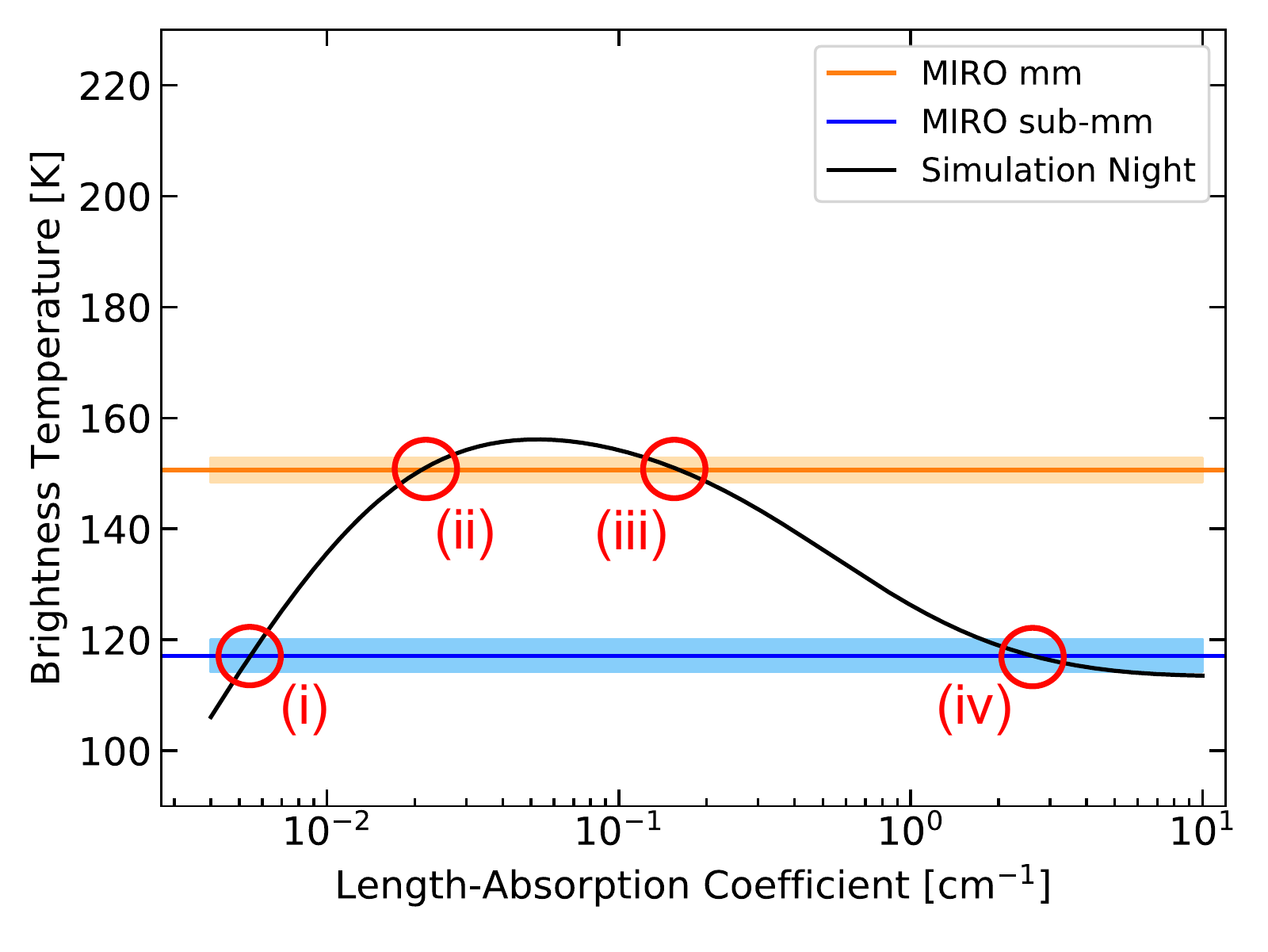}
	\end{center}	
	\caption[]{For the night cases, we found four solutions that are labeled with (i) - (iv). We excluded solutions (i) and (ii). Here, the Lambert-Beer results for the no-pebble case with a thermal conductivity of $\lambda = 4\cdot 10^{-3}\mathrm{\ Wm^{-1}K^{-1}}$ are shown.}
	\label{Fig:SimulatedTemp}
\end{figure}

\begin{figure}
	\begin{center}
		\includegraphics[width=1\columnwidth]{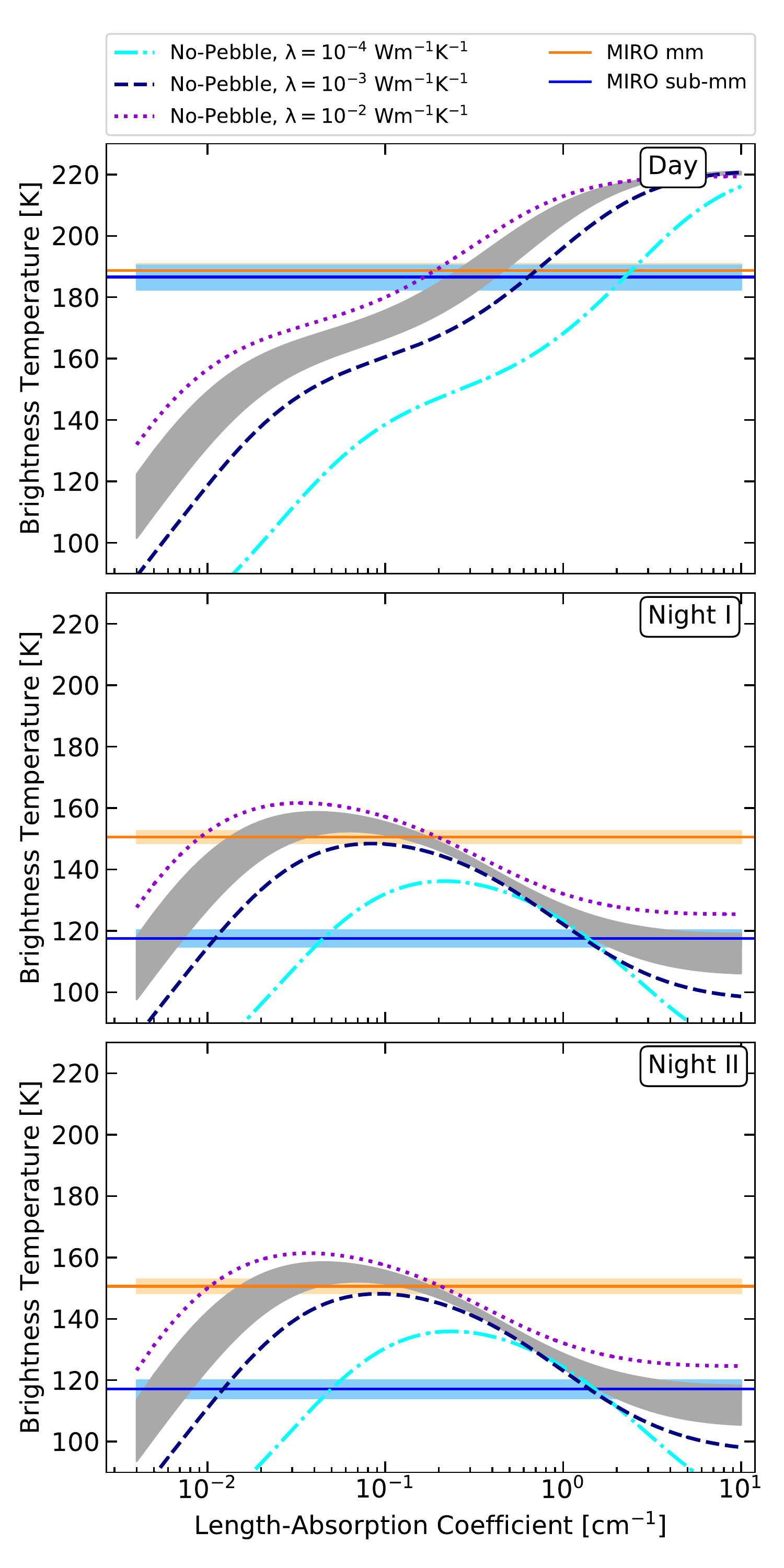}
		\caption{Overview of the simulated brightness temperatures derived from the Lambert-Beer law for the different cases investigated. The MIRO brightness temperatures are displayed in orange for the mm channel and in blue for the sub-mm channel. The lighter colours represent the respective uncertainties of the analysed MIRO brightness temperatures. The intersection marks the best-fitting length-absorption coefficient. In the no-pebble case, a solution for both channels in the night cases is only found for values of the thermal conductivity between $\lambda = 2\cdot 10^{-3}\mathrm{\ Wm^{-1}K^{-1}}$ and $\lambda = 6\cdot 10^{-3}\mathrm{\ Wm^{-1}K^{-1}}$ marked as the grey shaded area.}
		\label{Fig:SimulatedTB}
	\end{center}
\end{figure}

Fig.~\ref{Fig:SimulatedTemp} shows the simulated brightness temperatures derived by applying the Lambert-Beer law in black and the MIRO brightness temperatures in orange for the mm wavelength band and in blue for the sub-mm channel. The intersection marks the best-fitting length-absorption coefficient that is plotted on the logarithmic x-axis. In general, for high absorption coefficients the simulated brightness temperature saturates and converges with the surface temperature, while for very small absorption coefficients, the brightness temperature decreases towards the interior temperature of 50 K. In the night cases, there are two solutions for each channel, due to the temperature inversion shown in the derived temperature profiles (Fig.~\ref{Fig:TemperatureProfiles_nopebble}). The first solution (i) in Fig.~\ref{Fig:SimulatedTemp} for $\alpha_{\mathrm{sub-mm}}$ can be excluded, because it is lower than the values for $\alpha_{\mathrm{mm}}$. In general, $\alpha_{\mathrm{mm}} < \alpha_{\mathrm{sub-mm}}$ must be assumed, because no other behaviour has been measured in any known material. Moreover, we exclude the solution (ii) for $\alpha_{\mathrm{mm}}$ as well, as it is not consistent with the observed day-night temperature variations. Fig. \ref{Fig:SimulatedTB} provides an overview of all the simulated brightness temperatures in the no-pebble case. 

\subsection{\label{AP:RT}Ray-tracing approach}
\begin{figure*}
    \centering
    \includegraphics[width=\textwidth]{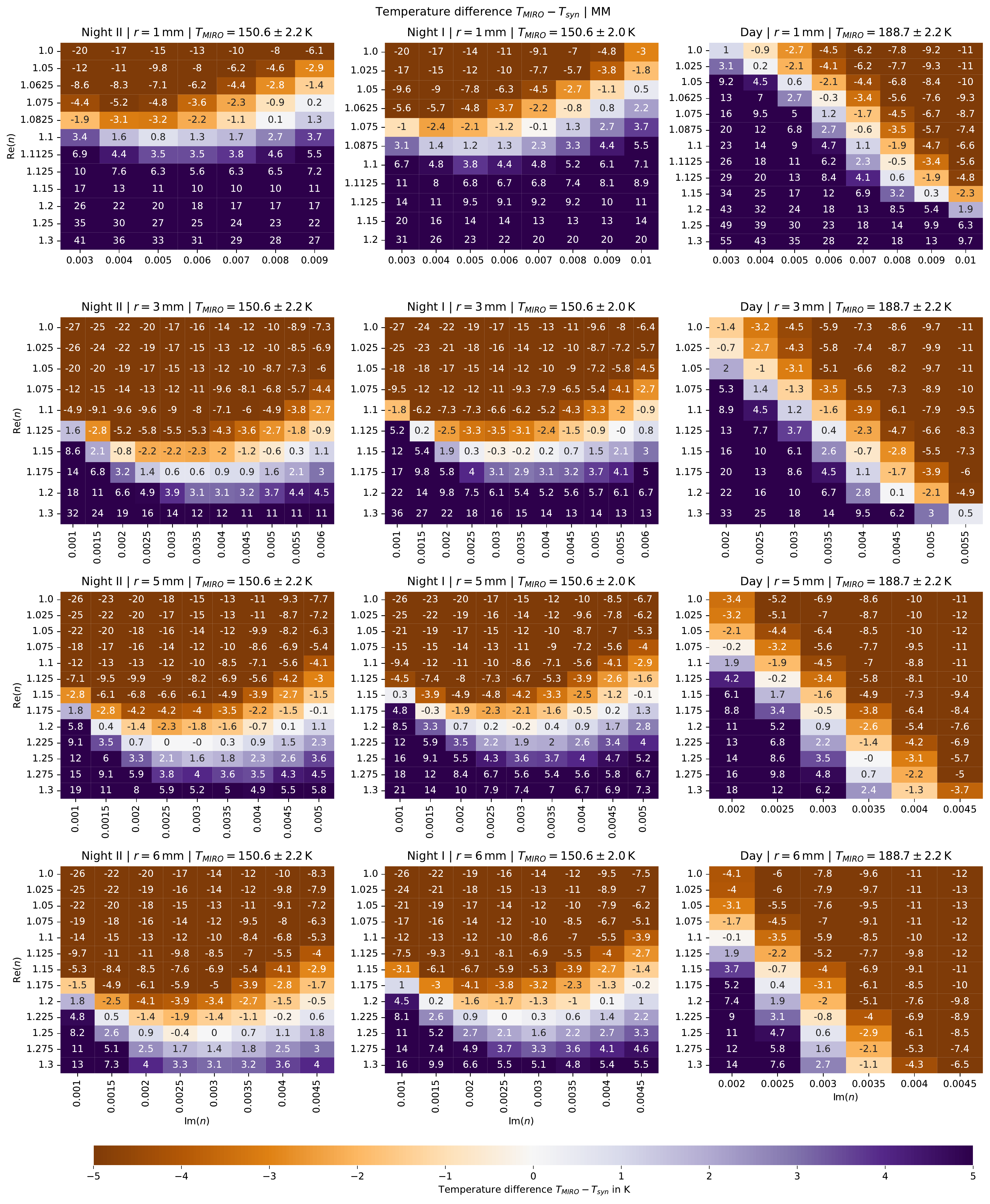}
    \caption{Temperature difference between the ray-tracing model with a pebble refractive index $n$ (see Sect.~\ref{sect:RTM}) and the MIRO measurement at mm wavelength for different pebble radii $r$ and observation cases.}
        \label{fig:RT_heatmap_MM}
\end{figure*}

\begin{figure*}
    \centering
    \includegraphics[width=\textwidth]{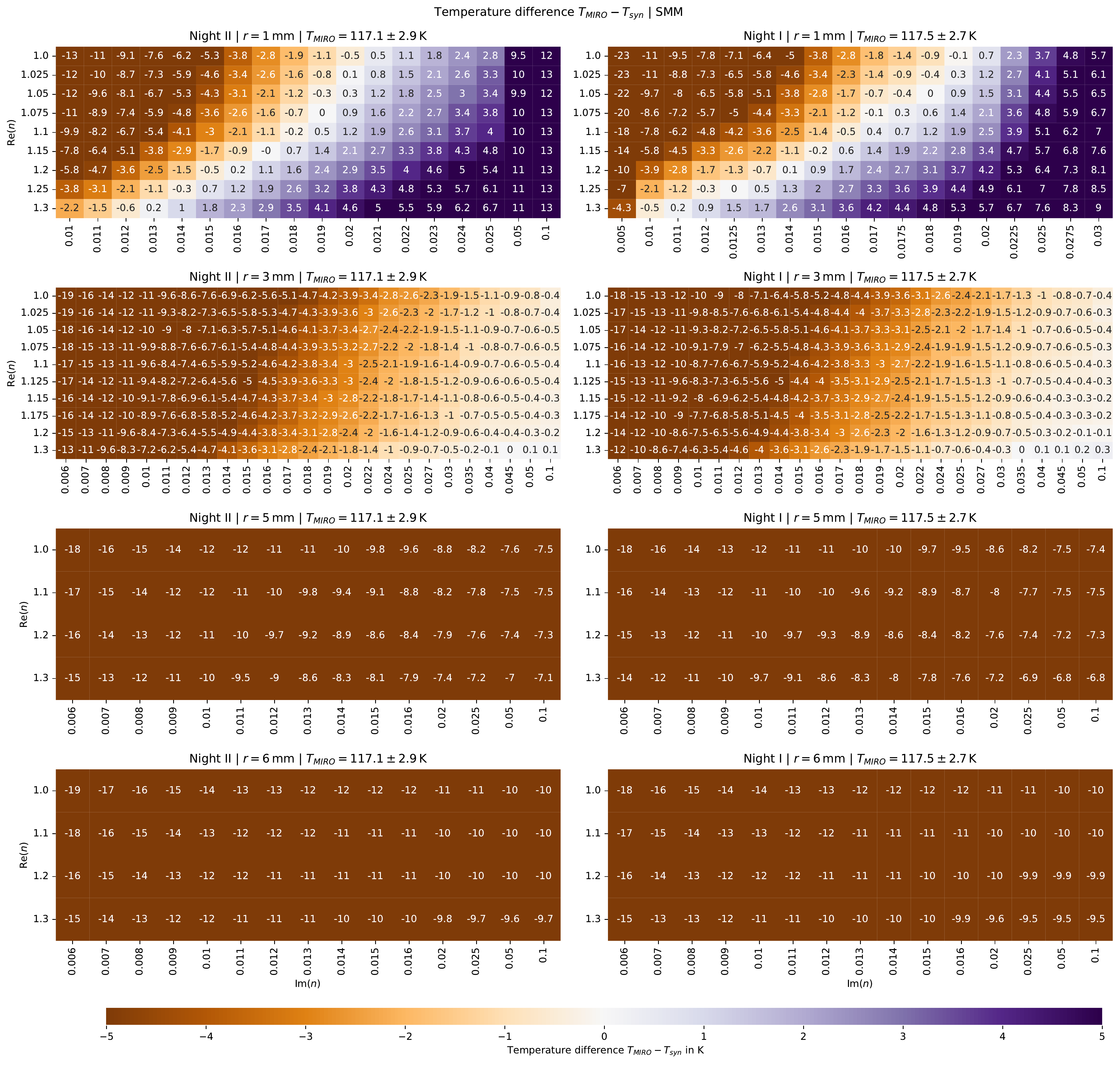}
    \caption{Temperature difference between ray-tracing model with a pebble refractive index $n$ (see Sect.~\ref{sect:RTM}) and the MIRO measurement at sub-mm wavelength for different pebble radii $r$ and observation cases.}
    \label{fig:RT_heatmap_SMM}
\end{figure*}

Figs.~\ref{fig:RT_heatmap_MM} and \ref{fig:RT_heatmap_SMM} show the temperature differences between the ray-tracing model with a pebble refractive index $n$ (see Sect.~\ref{sect:RTM}) and the MIRO measurement at mm wavelengths (Fig.~\ref{fig:RT_heatmap_MM}) and sub-mm wavelengths (Fig.~\ref{fig:RT_heatmap_SMM}) for different pebble radii $r$ and observation cases.

\section{\label{AP:WO}Waveoptics simulations}

\begin{figure}
	\begin{center}
		\includegraphics[width=0.6\columnwidth]{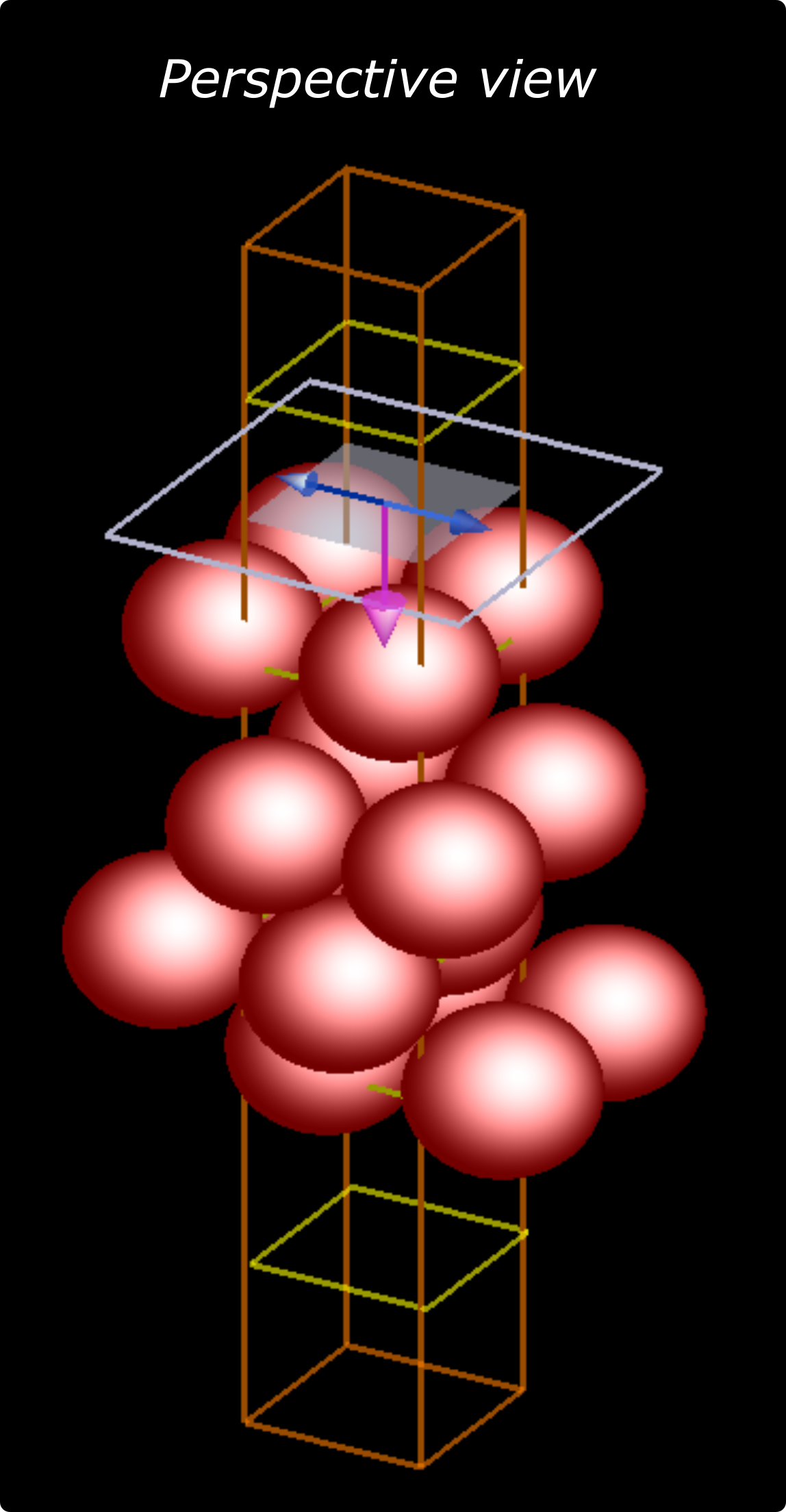}
		\caption{Illustration of the setup for the waveoptics simulations showing the pebble configuration, the start plane of the incident wave (grey frame) with propagation (purple arrow) and polarisation (blue double arrow) directions of the incident wave, the intensity monitor planes for reflection and transmission (yellow frames) and the border (orange) of the spatial region that is periodically repeated in the two lateral directions in order to create an infinite layer structure. The figure is a screenshot from the control screen of the FDTD package. }
		\label{Lumerical}
	\end{center}
\end{figure}

As ray-tracing simulations naturally ignore a number of effects related to the phase of the propagating wave, we performed additional simulations with both the ray-tracing and waveoptics methods, in order to study the influence of these effects on absorption and reflection within a pebble structure. For this purpose, we used a simplified geometry produced by periodically repeating a twisted chain of four pebbles in both the lateral directions, allowing the chains to touch so that a dense structure (filling factor 0.6) of four layers of pebbles was produced (see  Fig.\,\ref{Lumerical}). The simulations were carried out only at the mm wavelength, and for $r~=~1$~mm, 3~mm, and 6~mm pebbles, respectively. The incident radiation propagates along the axis of the twisted pebble chain, i.e., normal to the layer structure produced by the periodic repetition. The complex refractive index of the pebbles was only varied in its real part (n=1.08-1.25), while we fixed the imaginary part at a representative value of k=0.003. 

For the waveoptics simulations, we used both the commercial FDTD (finite-difference time-domain) package LUMERICAL\footnote{https://support.lumerical.com/} and the DDA method (discrete dipole approximation) implemented in the code DDSCAT \citep[Vers. 7.3.3.,][]{DDSCAT1994,DDAperiodic2008}. FDTD solves the Maxwell equations in time domain, based on a discretisation in space. Fig.\,\ref{Lumerical} illustrates the pebble structure (red), the so-called ``total-field, scattered-field plane wave light source'' (grey frame) with the propagation and polarisation directions, and the intensity monitor planes (yellow). The pebble structure within the orange frame is periodically repeated in the lateral directions. The spatial resolution is 0.04~mm in all simulations, i.e. 1/40 of the wavelength. The whole structure is enclosed by a perfectly matched layer (the ``PML-box'', not shown), which allows the light to leave the simulation volume with minimal reflection. DDSCAT was mainly used for small particle radii, but provided also results for $r~=~6$~mm pebbles and refractive indices above 1.2. DDA calculates the response of an array of polarisable point dipoles on a grid of certain resolution, which represent the object in space. In our simulation, the grid resolution was $\sim 3\%$ of the pebble radius, which is $D=0.2$~mm for the largest pebble size and 0.04~mm for the smallest. With $|n|=1.25$, the condition $|n|D<\lambda/2/\pi$ for validity of the DDA is just fulfilled. 

\begin{figure}
	\begin{center}
		\includegraphics[width=1\columnwidth]{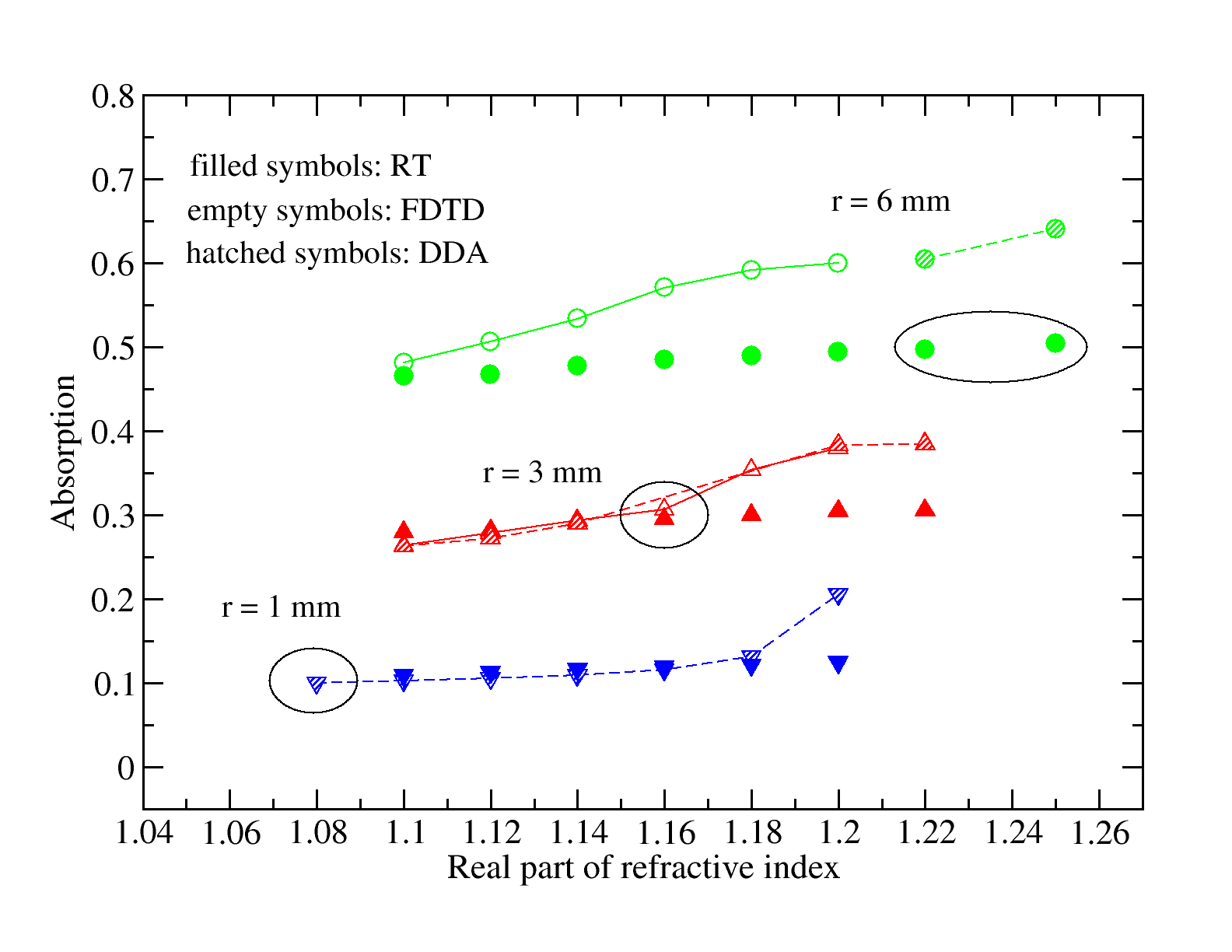}
		\caption{Comparison of absorbed intensities for wave propagation ($\lambda=1.6$ mm) through the pebble structure of Fig.\,\ref{Lumerical} at different real parts of the pebble refractive index. The imaginary part of the pebble refractive index has been set to $k$=0.003, the calculations have been performed for pebble radii of 1~mm, 3~mm, and 6~mm. Data shown as open and hatched symbols (solid and dashed lines) have been calculated with the FDTD and DDA waveoptics approaches, respectively, data represented by filled symbols are ray-tracing results. The ellipses indicate the approximate x-axis values at which solutions have been found in the modelling of the MIRO brightness temperatures. }
		\label{Wave_RA_coeff}
	\end{center}
\end{figure}

The results of the simulations, compared to those obtained with the ray-tracing method for the same geometry, are shown in Fig.\,\ref{Wave_RA_coeff}. Although both the intensities reflected from and absorbed within the pebble structure were calculated in each case, we show only the absorbed fraction of intensity. For the reflected intensity we found that it was heavily influenced by resonance effects of the wavelength with both the pebble size and the pebble layer thickness. In case of the small pebbles and small refractive indices, for instance, reflectivities varied between almost zero and the ray-tracing result. A detailed study of these effects was beyond our capabilities and also not useful, because the layer structure must be considered artificially produced by the construction of the pebble target. Thus, we restricted ourselves to the trends in the absorbed intensities, which show a much smaller dependency on the pebble size, especially for larger pebbles. 

The main result of the comparison are significant enhancements of the absorption occurring preferentially at higher refractive indices. 
For the $r=1$~mm pebbles, this effect sets in at $n=1.2$ and for larger pebble sizes at successively smaller refractive indices. Only at $r=6$~mm, this enhancement seems to be relevant for the solution of the brightness temperature simulation, as these required a refractive index above $n=1.2$ (encircled data points in Fig.\,\ref{Wave_RA_coeff}). It is argued that the enhancement effect occurs predominantly where the optical diameter of a pebble $2nr$ corresponds approximately to an odd multiple of $\lambda/2$, i.e., at $n=1.2$ for $r=1$~mm,~$3$~mm; at $n=1.133$ and $n=1.267$ for $r=6$~mm. For larger pebble sizes, the individual resonances, if present, would naturally be less well resolved than for smaller ones. If true, the absorption could be also enhanced at refractive index values relevant for smaller pebbles, if the sizes of these pebbles would just be slightly larger. Tests have shown, that for $n=1.08$ and $r=1.1$~mm this enhancement might still amount to about 10\%. The results of our waveoptics simulations are, unfortunately, not quantitatively applicable to the case of the MIRO observations, mainly because of the normal incidence of radiation that we had to apply. However, it should be considered to be likely, that waveoptics effects are able to enhance the average length-absorption coefficient in a pebble structure by factors of the order of 1.2-1.3, depending primarily on the value of the refractive index.   

\bsp	
\label{lastpage}
Figure\end{document}